\documentclass[twocolumn,aps,prx,showpacs,preprintnumbers,amsmath,amssymb]{revtex4-1}
\usepackage{natbib}
\usepackage{graphicx}
\usepackage{bm}
\usepackage{gensymb}
\usepackage{amsmath}
\usepackage{amssymb}
\usepackage{overpic}
\usepackage{color}
\begin{document}
\title{Study of decoherence in models for hard-core bosons coupled to optical phonons}
\author{A. Dey}
\author{M. Q. Lone} 
\author {S. Yarlagadda}
\affiliation{CMP Div.,
1/AF Salt Lake, Saha Institute of Nuclear physics, Kolkata, India -700064}
\date{\today}
\begin{abstract}
Understanding coherent dynamics of excitons, spins, or hard-core bosons (HCBs) has tremendous
scientific and technological implications for light harvesting  and quantum computation.
Here, we study decay of excited-state population and decoherence in two models for HCBs, 
 namely: an infinite-range HCB model governed by Markovian dynamics and a two-site HCB model with site-dependent strong potentials
and subject to non-Markovian dynamics. Both models are investigated in the regimes of antiadiabaticity and strong
HCB-phonon coupling with each site providing a different local optical phonon environment; furthermore,
 the HCB systems in both models are taken to be initially uncorrelated with the environment in the polaronic frame of reference.  
For the infinite-range model,
we derive an effective many-body Hamiltonian 
that commutes with the long-range system Hamiltonian
and thus has the same set of eigenstates; consequently, 
a quantum-master-equation approach 
shows that the quantum states of the system do not decohere.
In the case of the two-site HCB model, 
we show clearly that the degree of decoherence and decay of excited state 
are enhanced by the proximity of the site-energy difference to  
the eigenenergy of  phonons and are most pronounced when the site-energy difference is at resonance with
twice the polaronic energy; additionally, the decoherence and the decay  effects are reduced when the strength of HCB-phonon
coupling is increased.
Even for a multimode situation, the degree of decoherence and decay are again dictated
by the nearness of the energy difference to the allowed phonon mode eigenenergies. 
\end{abstract}
\maketitle
\section {Introduction}
Quantum information processing heavily relies on  a precious and fragile 
resource, namely,  
quantum entanglement \cite{6}.
The  fragility
of entanglement is due to the coupling between a quantum system and its environment;
such a coupling leads to decoherence, the process
 by which information is degraded.
Decoherence is  the fundamental mechanism by which fragile superposition of states is destroyed
thereby producing 
a quantum to 
classical transition \cite{schloss,zurek2}. Since coupling of a quantum system
to the environment and the concomitant entanglement fragility are 
ubiquitous
 \cite{6,schloss}, it is imperative that progress be made in
minimizing decoherence.

In general, a many-qubit (i.e., many-particle) system can have 
distance-dependent interaction. The two limiting cases for interaction are
particle 
(HCB) hopping strength that is independent of distance and a system with nearest-neighbor
hopping only.
{In this work, we consider
an extreme model involving distance-independent interaction of HCBs which can be mapped onto the following spin-model: 
$ \sum_{i,j>i} [-J_{\perp}({S_i^{x}}{S_j^{x}} + {S_i^{y}}{S_j^{y}}) 
+ J_{\parallel}  S_i^z S_j^{z}] $. 
Such a model has relevance to many realistic physical systems of interest.
Firstly, the well-studied Lipkin-Meshkov-Glick (LMG) model \cite{lmg}
$H_{\rm LMG} = -2h(\sum_jS_j^z) -2\lambda[(\sum_j S_j^x)^2+\gamma (\sum_jS_j^y)^2]/N$͒
(for $h=0$ and $\gamma=1$) is a special case of the above mentioned long-range model (for $J_{\parallel}=0$ and $J_{\perp} =\lambda$);
while, for $h=0$ and $\gamma=0$, LMG model is a special case of the model for $J_{\perp}=0$ and $J_{\parallel}= -\lambda$. 
{ Long-range interactions can actually occur quite naturally in cavity quantum electrodynamics}; by varying the external model parameters, 
it has been proposed
that positive and negative values of $\lambda$ as well as $-1 \le \gamma \le 1$ values can be achieved \cite{parkins}.
Secondly, it has been shown by Ezawa that the long-range ferromagnetic Heisenberg model describes
well a zigzag graphene nanodisc \cite{ezawa}.  A two-spin system and a four-spin system (with spins at the corners of
a regular tetrahedron) can be physically realized (for instance from a Hubbard model)
as exact special cases of the long-range model; it is conceivable that slightly larger
clusters of particles (for instance, clusters containing 6 or 8 particles) can be physically realized as reasonable approximations of such a model. 

{ Highly efficient coherent energy transfer processes in light-harvesting complexes is an active area of research \cite{engel}. 
Two-dimensional electronic spectroscopy gives a picture of the evolution of the
density matrix and enables mapping of populations and coherences \cite{Fleming}.
Fully connected network (FCN) is a well-studied model in the context of excitation energy transfer (EET) in 
Fenna-Matthews-Olson (FMO) complexes 
\cite{Alex1}. FCN is characterized by uniform hopping strength between any pair of sites 
(chromophores in the case of FMO complexes) 
 and is an extreme limit of long-range interaction model for excitons. Moreover, 
the phonon fluctuations at various sites (chromophores) are uncorrelated to each other \cite{Fleming}, i.e., local phonon 
effects are significant in such complexes. The system-bath coupling in photosynthetic complexes is thought to be 
not weak but to be at least in the intermediate regime 
\cite{Fleming}; 
instead of employing the usual quantum master equation techniques valid for the weak-coupling limit,
 modified approaches valid for  broader range of couplings have been studied.} 

{Modelling and controlling the environment of a solid-state quantum bit is a major challenge
in quantum computation. Fairly long coherence times have been achieved
in semiconductor-based double quantum dots where the qubit information is encoded
in the singlet-triplet states of two spins with total $S^z$ equal to zero \cite{semidqd}. In these quantum dot systems, 
spin states are prepared, manipulated, and measured using rapid control of Heisenberg exchange interaction. However, 
the small size of a semiconductor qubit is limited by the extent of the electronic wavefunction which is more than a few nanometers.} 

{In this work, without specifically modelling either a light-harvesting system or a qubit device,
 we would like to identify and understand some
important features of related HCB systems which lead to
 either insignificant or sufficiently weak decoherence and decay of excited-state population.
To this end, in the regimes of strong HCB-phonon interaction and antiadiabaticity, we study two HCB models in the polaronic frame of reference
where initially
the system and the environment constitute a simply separable
state. In the transformed polaronic frame, the interaction term is weak and
enables use of perturbation theory; furthermore, both preparation and measurement can be done in the dressed (polaronic)
basis \cite{lidar2,polarondyn}.
We first analyze 
the nature of 
decoherence and decay of excited-state population in an infinite-range interaction model for HCBs
that are coupled to local optical phonons.
We show that the effective
Hamiltonian in second-order
  perturbation theory
retains the same eigenstates as
the infinite-range system Hamiltonian. Our dynamical analysis 
 shows that the system,  
 when Markov processes are considered, neither decoheres
nor allows decay of excited states.   
Next, for the more realistic situation where
lattice sites have different site-energies and the dynamics is non-Markovian,
(instead of a many-body problem) we analyze a more tractable case where just 
one HCB is hopping between two sites 
 and the HCB-phonon coupling is local. Using non-Markovian second-order quantum master equation,  we find
that decoherence as well as decay of the population of the excited state are small if the site-energy difference is sufficiently different
from the phonon eigenenergies; these features are manifested  for both single-mode and multimode optical phonons.
 }

The rest of the paper is organized as follows.
In Sec. II, we introduce the infinite-range HCB Hamiltonian 
strongly coupled to 
local optical phonons and
derive the effective Hamiltonian.  
In Sec. III, 
 using the master-equation approach, 
we study decoherence under Markovian dynamics. 
  Next, in Sec. IV, we study decoherence and decay of excited state 
using a non-Markovian analysis
for a system of two sites (each with a different site energy).
 Finally, in Sec. V, we give our conclusions and make some general remarks
regarding the wider context of our results.
{The paper also has an appendix containing  detailed calculations for the terms of the master equation 
used for the two-site case.

\section{Infinite-Range HCB Model 
with HCBs coupled to local optical phonons}
 We begin by introducing the infinite-range HCB model whose decoherence will be studied
when the system is coupled to
 local phonons.
The Hamiltonian for that is defined as:
\begin{eqnarray}
H_{\rm HCB}&=& \sum_{i,j>i}[\frac{-J_{\perp}}{2}( b^{\dagger}_i  b_{j} +{\rm H.c.}) + J_{\lVert} (n_i - \frac{1}{2})(n_j -\frac{1}{2}) ] ,\nonumber \\
\label{ir}
\end{eqnarray}
where $\frac{J_{\perp}}{2}$ and  $J_{\lVert}$ ($J_{\perp}>0$ and $J_{\lVert}>0$) are the hopping and HCB repulsion 
strengths between different sites, respectively. 
HCB creation and destruction operators are defined as $b^{\dagger}_i$ and $b_i$ with the commutation relations given by
\begin{eqnarray}
[b_i,b_j]&=&[b_i,b^{\dagger}_j]= 0 , \textrm{ for } i \neq j , \nonumber\\
\{b_i,b^{\dagger}_i\}& = & 1  ,
\label{commute}
\end{eqnarray}   
and $n_i\equiv b^{\dagger}_i b_i$. In Eq. (\ref{ir}), it is understood that $J_{\perp} = J^{\star}_{\perp}/(N-1)$ and 
$J_{\lVert} = J^{\star}_{\lVert}/(N-1)$
(where $ J_{\perp}^{\star}$ and $ J_{\lVert}^{\star}$ are finite
quantities) so that the energy per site remains finite as $N \rightarrow \infty$.
The total Hamiltonian is defined by
\begin{eqnarray}
H_T &=&   \sum_{i,j>i}[\frac{-J_{\perp}}{2}(b^{\dagger}_i  b_{j} +{\rm H.c.}) + J_{\lVert} (n_i - \frac{1}{2})(n_j -\frac{1}{2})  ] 
\nonumber \\
          &&+ \omega \sum_j a^{\dagger}_{j} a_j
          + g \omega \sum_j (n_j-\frac{1}{2}) 
 (a_j +a^{\dagger}_j) , 
 \label{Ham3}
\end{eqnarray}
where $a_j$ and $a^{\dagger}_j$ are the destruction and creation operators of phonons, respectively, $g$ is the HCB-phonon coupling constant,
 and $\omega$ is the phonon frequency. 
Subsequently, we perform 
the well-known Lang-Firsov (LF) transformation \cite {lang,sdadys} on this
Hamiltonian. 
Under the LF transformation  given by 
\begin{eqnarray}
{{H^{ L}_T}} \equiv e^S H_T e^{-S} =H^{L}_0+H^{L}_I ,
\end{eqnarray}
 with
$S= - g \sum_i (n_i-\frac{1}{2}) (a_i - a^{\dagger}_i)$, the operators
$b_j$ and $a_j$ transform like fermions and bosons. 
Next, the unperturbed Hamiltonian $H_0^L$ is expressed as 
\cite{sdadys}
\begin{eqnarray}
\!\!
H^{L}_0 =  H^{L}_s +H^{L}_{env} ,
\label{H0}
\end{eqnarray}
where  we identify $H^{L}_s$ as the system Hamiltonian
\begin{eqnarray}
\!\!\!\! H^{L}_s &=&  \sum_{i,j>i}[\frac{-J_{\perp}}{2} e^{-g^2}(b^{\dagger}_i  b_{j} +{\rm H.c.})
\nonumber \\
&& ~~~~~~~~~~ + J_{\lVert}  (n_i -\frac{1}{2})( n_j - \frac{1}{2})  ] ,
\label{Hs}
\end{eqnarray}
and $H^{L}_{env}$ as the Hamiltonian of the environment
\begin{eqnarray}
H^{L}_{env} = \omega \sum_j a^{\dagger}_j a_j . 
\label{Henv}
\end{eqnarray}
On the other hand, the interaction $H^{L}_I$ which we will treat as perturbation is given by
\begin{eqnarray}
H^{L}_I
= \frac{-J_{\perp}}{2}e^{-g^2} \sum_{i,j>i}[b^{\dagger}_i  b_{j} ]
            \{\mathcal S^{{ij}^\dagger}_+ \mathcal S^{ij}_{-}-1\} +{\rm H.c.} ,
\label{int}
\end{eqnarray}
where $\mathcal S^{ij}_{\pm} = \textrm{exp}[\pm g(a_i - a_{j})]$. 
 In the transformed frame, the system Hamiltonian $H_s^L$ depicts that all the HCBs are coupled to the same
phononic mean-field. Thus, the unperturbed Hamiltonian $H^{L}_0$ comprises of
the system Hamiltonian $H^{L}_s$ representing
  HCBs with the same reduced hopping term $\frac{1}{2} J_{\perp} e^{-g^2}$
and the environment Hamiltonian $H^{L}_{env}$ involving displaced bath oscillators corresponding to local distortions.
Here it should be pointed out that the mean-field term $H_s^L$ 
  involves controlled degrees of freedom. 
  Thus  no irreversibility
is involved 
under evolution due to $H^{L}_0$.
 {\it On the other hand, perturbation $H^{L}_{I}$ pertains to the interaction of HCBs with local 
deviations from the phononic mean-field;
  the interaction term $H^{L}_I$ represents numerous
or uncontrolled environmental degrees of freedom and thus has the potential
for producing decoherence}. Furthermore, it is of interest to note that
the interaction term is weak in the transformed frame
{unlike the interaction 
in the original frame}; thus one can perform perturbation theory with the interaction term.
We represent the  eigenstates of the unperturbed Hamiltonian $H^{L}_{0}$ as
$|n,m\rangle\equiv|n\rangle_{s}\otimes|m\rangle_{ph}$ with the corresponding
 eigenenergies $E_{n,m}=E_{n}^{s}+E_{m}^{ph}$;
$|n\rangle_{s}$ is the eigenstate of the system with eigenenergy $E_{n}^{s}$
while $|m\rangle_{ph}$ is the eigenstate for  the environment with eigenenergy $E_{m}^{ph}$.
 Henceforth, for brevity, we will use $\omega_m \equiv E_m^{ph}$.
On observing that $\langle0,0|H^{L}_{I}|0,0\rangle=0$ ({ i.e., the ground state
expectation value of the deviations is zero}),
we obtain the next relevant second-order perturbation term \cite{sdadys}
\begin{eqnarray}
E^{(2)}=\sum_{n,m}{{\langle0,0|H^{L}_{I}|n,m\rangle\langle n,m|H^{L}_{I}|0,0\rangle}\over{E_{0,0}-E_{n,m}}} .
\label{E2}
\end{eqnarray}
{For strong coupling {{(i.e., $g^2 \gg 1$) \cite{g2} and non-adiabatic (i.e., $J_{\perp}^{\star}/\omega \leq 1$) \cite{c60} conditions
assumed in this paper, it follows that $J_{\perp}^{\star}e^{-g^2} \ll \omega$.
On noting that 
$\omega_{m}-\omega_{0} =\omega_m$ is a positive integral multiple of $\omega$  and that
$E_{n}^{s}-E_{0}^{s}
\le J_{\perp}^{\star}e^{-g^2} $ (as shown in the next section)}}, we 
get the following second-order term $H^{(2)}$ \cite{sdys}
using Schrieffer-Wolff (SW) transformation (as elaborated in
Appendix A of Refs. \onlinecite{srsypbl} and \onlinecite{srsypbl2}):}
\begin{eqnarray}
\!\!\!\!H^{(2)} \! &=& \!
-\sum_{m}{{{_{ph}\langle0|H_{I}|m\rangle_{ph}}~{_{ph}\langle m|H_{I}|0\rangle_{ph}}}\over{\omega_{m}}} 
\nonumber \\
&=& \! \sum_{i, j > i } 
 \left [\Big(\frac{1}{2} J_{\perp}^{(2)}
b^{\dagger}_i
 b_j +{\rm H.c.}\Big) \right .
\nonumber \\
&&~~~~~ \left . - 
 \frac{1}{2} J_{\parallel}^{(2)}
 \{n_i(1-n_j)+n_j(1-n_i)\} \right ] ,
\label{H_eff}
\end{eqnarray}
where 
\begin{eqnarray}
\!\!\!\!\!\!\!\!\!\!\!\! J_{\perp}^{(2)} \equiv  - (N-2) f_1 (g) \frac{J_{\perp}^2 e^{-2g^2}}{2 \omega}
 \sim -(N-2) \frac{J_{\perp}^2e^{-g^2}}{2g^2 \omega} ,
\label{j_perp}
 \end{eqnarray}
\begin{eqnarray}
 J_{\parallel}^{(2)} \equiv 
[2f_1 (g)+f_2(g)]\frac{J_{\perp}^2 e^{-2g^2}}{{2 \omega}} 
\sim \frac{J_{\perp}^2 }{{4g^2 \omega}}, 
\label{j_para} 
\end{eqnarray}
 with
$f_1(g) \equiv \sum^{\infty}_{l=1} g^{2l}/(l!l)$
 and
$f_2(g) \equiv \sum^{\infty}_{j=1}\sum^{\infty}_{l=1} g^{2(j+l)}/[j!l!(j+l)]$. 
{The effective Hamiltonian $H^{L}_{s}+H^{(2)}$ is a low energy Hamiltonian
obtained by the canonical SW transformation \cite{schrieffer,loss2} decoupling the low-energy and the
high-energy subspaces; this decoupling  is a consequence of $J_{\perp}^{\star}e^{-g^2} \ll \omega$. }
We now make the important observation that 
the effective Hamiltonian $H^{L}_{s}+H^{(2)}$ has the same set of eigenstates as those of $H^{L}_s$ and $H_{\rm HCB}$ 
because $\sum_{i, j > i } 
 (n_i-\frac{1}{2})(n_j-\frac{1}{2})$ commutes with both $H_s^L$ and $ H_{\rm HCB}$. Actually, we have shown that even in higher-order perturbation theory (higher than second order) 
 the eigenstates of the effective Hamiltonian do not change \cite{admlsy}. 
 The small parameter of our perturbation theory, for a small $N$ system, is $\frac{J_{\perp}}{g \omega}$ (see Ref. \onlinecite{sy1} for details);
whereas for a large $N$, the small parameter is $\frac{J_{\perp}^{\star}}{g^2 \omega}$ (see Ref. \onlinecite{sm_par} for an explanation).
It is the long range of the model that enables  
 the eigenstates of the system to 
remain unchanged.
{While the fact that the eigenstates of the effective Hamiltonian remain
the same as those of $H_{\rm HCB}$ may be suggestive of the robustness of the states of this long-range model,
to establish that the states of the system are actually decoherence free, it is necessary to show that the off-diagonal 
  matrix elements of the
system's reduced density matrix do not diminish.}
Next, we study decoherence in a dynamical
context and gain more insight into how the states
of our $H_{\rm HCB}$ can be decoherence free.

{\section{Dynamical evolution of the system}}
In this section,
we will study decoherence in the system from the dynamical perspective.
We will discuss the dynamics of an
open quantum system, described by the $H_{\rm HCB}$, using master equation approach. 
Our quantum system is open because it is coupled to
another quantum system, i.e., a bath or environment \cite{Pet}. 
In our case, $H_{\rm HCB}$ is coupled to a bath of local optical phonons
{ [see Eq. (\ref{Ham3})].}
As a consequence of the system-environment  coupling, 
the state of the system  may change.  This interaction may lead to 
certain system-environment correlations such that
the resulting state of the system
may  no longer be represented in terms of unitary Hamiltonian dynamics. 
The dynamics of the system, described by the reduced density matrix $\rho_s(t)$ at 
time $t$, is obtained from  the density matrix $\rho_T(t)$ of the total system by taking the 
partial trace over the degrees of freedom of the environment:
\begin{eqnarray}
\rho_s(t) = Tr_R\left[\rho_T(t)  \right]= Tr_R\left[  U(t) \rho_T(0) U^{\dagger}(t)     \right] ,
\end{eqnarray}
where $U(t)$  represents the time-evolution operator of the total system. Now it is
evident from the above equation that we need first to determine 
the dynamics of the total system which is a  difficult task 
in most of the cases. 
By contrast, master equation approach 
conveniently  and directly yields the time evolution of
the reduced density matrix 
 of the system interacting  with an environment.
This approach relieves us from the need of having to first determine the dynamics of the
total system-environment combination and then to trace out the degrees
of freedom of the environment.\\

{We begin by observing that,
to understand decoherence in the original frame of reference where the HCB-phonon coupling is strong,
 it is convenient to use the LF transformed frame
of reference:
 in the LF frame the system-environment coupling is weak,
and furthermore, a polaron (represented by $e^{-S} b^{\dagger}|0\rangle_{s}\otimes|0\rangle_{ph}$) that is entangled
 with the environment in the original frame of reference becomes 
unentangled in the LF frame
(i.e., it becomes an undressed particle represented by
$ b^{\dagger}|0\rangle_{s}\otimes|0\rangle_{ph}$). The relevant Hamiltonian (for our decoherence analysis)
is the following LF transformed Hamiltonian:}
\begin{eqnarray}
H^{\rm L}_T=  H^{L}_0 + H^{L}_I ,
\end{eqnarray}
where $H^{L}_0$ is the system-environment Hamiltonian given by Eq. (\ref{H0}) 
and $H^{L}_I$ represents the interaction Hamiltonian given by Eq. (\ref{int}). 
It is convenient and simple to derive the quantum master equation
in the  interaction picture. Thus our starting point is the interaction picture
von Neumann equation for the total density operator $\tilde{\rho}_T(t)$ 
\begin{eqnarray}
\frac{d \tilde{\rho}_T (t)}{dt} = -i [\tilde{H}^{L}_I(t), \tilde{\rho}_T(t)] ,
\label{von1}
\end{eqnarray} 
where $\tilde{H}^{L}_I(t) = e^{iH^{L}_0t} H^{L}_I e^{-iH^{L}_0t} $ and 
$\tilde{\rho}_T (t) = e^{iH^{L}_0t} \rho_T(t) e^{-iH^{L}_0t} $ are the interaction
Hamiltonian and the total system density matrix operators (respectively) 
expressed in the interaction picture.
 Re-expressing the above equation in integral form yields
\begin{eqnarray}
\tilde{\rho}_T(t) = \tilde{\rho}_T(0) - i \int_{0}^t d \tau [\tilde{H}^{L}_I(\tau), \tilde{\rho}_T(\tau)].
\label{von2}
\end{eqnarray}

{{Nowadays there is considerable interest in systems with initial correlation with the environment \cite{modi,morozov}.
Here too the initial state of the 
total system, in the original untransformed frame of reference,
 is taken to be made up of particles entangled with the environment (i.e., polarons or particles
dressed with environmental phonons), 
so that in the LF frame the initial state transforms to a factorized (or simply separable) state given as
$ \rho_T(0) = \rho_s(0) \otimes R_0$  with  
{ $R_0=\sum_{n}|n \rangle_{ph}~\!\! _{ph}\langle n|e^{-\beta \omega_n}/Z$}
being the initial thermal density matrix operator of the environment
 and temperature being equal to $\frac{1}{k_B \beta}$; 
furthermore, 
$Z=\sum_n e^{-\beta \omega_n}$
 defines the partition function of the environment. 
Here it should be pointed out that states can be prepared and measured in the dressed
(polaronic) basis; the dressed (polaronic) basis can be used for input and output \cite{lidar2}. 
The possible preparation of this initial factorized state is discussed for a realistic system, i.e., the double quantum dot (DQD) 
in Ref. \onlinecite{polarondyn}. 
{Moreover, if the phonon deformation time scale is much smaller than the bare hopping time scale 
(which is justified in strong coupling and non-adiabatic limit), the initial separability condition is certainly achievable. More specifically, in such a limit, 
as soon as a particle is put at any site, the local phonons quickly respond and reorganize to a new equilibrium position; consequently, the 
separable initial state (in the polaronic frame of reference) is formed.}
With this assumed initial state, we 
substitute  Eq. (\ref{von2}) inside the commutator of
 Eq. (\ref{von1}) and  then take the trace over the environmental degrees 
of freedom to obtain the following equation}:
\begin{eqnarray}
\!\!\!\!\!\!\!\!\!\!
\frac{d \tilde{\rho}_s(t)}{dt} &=& 
-i~Tr_R[\tilde{H}^{L}_I(t), \tilde{\rho}_s(0) \otimes R_0] \nonumber \\
&-& \int_0^t d\tau Tr_R[\tilde{H}^{L}_I(t),[\tilde{H}^{L}_I(\tau), \tilde{\rho}_T(\tau)]] .
\label{mas1}
\end{eqnarray}

The 
above equation  still contains 
the total density matrix $\tilde{\rho}_T(\tau)$; in order to  evaluate
it, we rely on an approximation known as the Born approximation. This approximation
assumes that the environment degrees of freedom are large and thus the effect on
the environment due to the system is negligibly small for a weak system-environment coupling.
As a consequence, we write   $\tilde{\rho}_T(\tau) = \tilde{\rho}_s(\tau) 
\otimes R_0 + \mathcal{O}(\tilde{H_I})$ 
within the second-order perturbation in system-environment interaction \cite{Pet,HJ,CM,YJ1,YJ2,HFPB,HPB,MS,EF,goan}.
 Therefore we can write  Eq. (\ref{mas1}) in time-local form as
\begin{eqnarray}
\!\!\!\!\!\!\!\!\!\!
\frac{d \tilde{\rho}_s(t)}{dt} &=& 
-i~Tr_R[\tilde{H}^{L}_I(t), \rho_s(0) \otimes R_0] \nonumber \\
&-& \int_0^t d\tau Tr_R[\tilde{H}^{L}_I(t),[\tilde{H}^{L}_I(\tau), \tilde{\rho}_s(t) \otimes R_0]] .
\label{mas2}
\end{eqnarray}
We note here that, for obtaining the non-Markovian time-convolutionless  master equation (\ref{mas2}),
we replaced $\tilde{\rho}_s(\tau) $ with $\tilde{\rho}_s(t)$. This replacement
is equivalent to obtaining a time-convolutionless master equation perturbatively 
up to only second order in the interaction Hamiltonian using the 
time-convolutionless projection operator technique \cite{Pet, HFPB, HPB}.
It has been shown in a number of cases 
that time-local approach works better than
time-nonlocal approach \cite{Pet, YJ1, MS, EF, UK}.
{ Now we will consider the 
second-order, time-convolutionless master equation (\ref{mas2}) with
the time variable $\tau$ replaced 
by $(t - \tau)$.
\begin{eqnarray}
\!\!\!\!\!\!\!\!\!\!
\frac{d \tilde{\rho}_s(t)}{dt} &=& 
-i~Tr_R[\tilde{H}^{L}_I(t), \rho_s(0) \otimes R_0] \nonumber \\
&-& \int_0^t d\tau Tr_R[\tilde{H}^{L}_I(t),[\tilde{H}^{L}_I(t-\tau), \tilde{\rho}_s(t) \otimes R_0]] .
\label{mas3}
\end{eqnarray}

 Next, we will study the Markovian dynamics of the system. To this end we assume that the correlation 
time scale  $\tau_{c}$  for the environmental fluctuations 
is negligibly small compared to the relaxation time scale $\tau_{s}$  for the system,
i.e.,  $\tau_{c} \ll \tau_{s}$. This time scale assumption is 
 {motivated}
by the condition 
 $J^{\star}e^{-g^2} \ll \omega$ already mentioned in Sec. $2$. 
The Markov approximation ($\tau_{c} \ll \tau_{s}$)
allows us to set the upper limit of the integral to $\infty$ in Eq. (\ref{mas3}).
 Thus we obtain the second-order time-convolutionless
 Markovian master equation 
 { (see Ref. \onlinecite{Pet} for further details)}:}
\begin{eqnarray}
\!\!\!\!\!\!\!\!\!\!
\frac{d \tilde{\rho}_s(t)}{dt} &=& 
-i~Tr_R[\tilde{H}^{L}_I(t), \rho_s(0) \otimes R_0] \nonumber \\
&-& \int_0^\infty d\tau Tr_R[\tilde{H}^{L}_I(t),[\tilde{H}^{L}_I(t - \tau), \tilde{\rho}_s(t) \otimes R_0]] .
\label{mark}
\end{eqnarray}
Defining $\{|n\rangle_{ph}\}$ as the basis set for phonons, therefore, we can write
the master equation as:
\begin{widetext}
{\begin{eqnarray}
\frac{d \tilde{\rho}_s(t)}{dt} &=&-i  \sum_{n}~ _{ph}\langle n| [\tilde{H}^{L}_I(t), \rho_s(0) \otimes R_0] |n \rangle_{ph}
\nonumber \\
&& 
- \sum_n \int_0^\infty d\tau \Big[~ _{ph}\langle n |
 \tilde{H}^{L}_I(t)\tilde{H}^{L}_I(t-\tau)\tilde{\rho}_s(t) \otimes R_0|n\rangle_{ph}  
 - ~_{ph}\langle n| \tilde{H}^{L}_I(t)\tilde{\rho}_s(t) \otimes R_0 \tilde{H}^{L}_I(t-\tau) |n\rangle_{ph} 
\nonumber \\
&&
 ~~~~~~~~~~~~~~~~~~ 
- ~_{ph}\langle n| \tilde{H}^{L}_I(t-\tau)\tilde{\rho}_s(t) \otimes R_0 \tilde{H}^{L}_I(t) |n\rangle_{ph}
+ ~_{ph}\langle n | \tilde{\rho}_s(t) \otimes R_0 \tilde{H}^{L}_I(t-\tau)\tilde{H}^{L}_I(t)|n\rangle_{ph}
 \Big] .
\label{mas5}
\end{eqnarray}}
\end{widetext}
In order to simplify the above master equation, we need to evaluate  
the time evolution of the operators involved in $H_I^L$. Considering the 
second term in Eq. (\ref{mas5}) yields
{\begin{eqnarray}
&& \!\!\!\!\!\!\! _{ph}\langle n |
 \tilde{H}^{L}_I(t)\tilde{H}^{L}_I(t-\tau)\tilde{\rho}_s(t) \otimes R_0 |n\rangle_{ph} 
\nonumber \\
&& \!\!\!\!\!\!\!=\sum_m e^{iH^{L}_s t} {_{ph}\langle n | H^{L}_I |m\rangle_{ph}} e^{-iH^{L}_s t} ~
    e^{iH^{L}_s (t-\tau)} \nonumber \\
 && \!\!\!\!\!\!\! \times{_{ph}\langle m| H^{L}_I|n\rangle_{ph}} e^{-iH^{L}_s( t-\tau)} \tilde{\rho}_s(t)
\frac{e^{-\beta \omega_n}}{Z} e^{i(\omega_n-\omega_m)\tau} .
\label{time}
\end{eqnarray}}

We connect the HCBs in real space with those in momentum space as:
$ b^{\dagger}_j =
 \frac{1}{\sqrt{N}}\sum_{k_1} e^{ik_1r_j }~ b^{\dagger}_{k_1}$ and $ b_j = \frac{1}{\sqrt{N}}\sum_{k_1} e^{-ik_1r_j }~ b_{k_1}$;
{{henceforth, in momentum space, the creation and destruction operators of HCBs
shall be denoted, respectively, as $b^{\dagger}_{k_n}$
and $b_{k_n}$ with $n=1,2, 3,...$.}}
Then, it is important to note that the hopping term 
in the system Hamiltonian can be written as 
{{(see Refs. \onlinecite{sdys,roth})}}:
\begin{eqnarray}
\frac{1}{2} J_{\perp}e^{-g^2} \sum_{i,j>i}( b^{\dagger}_i  b_{j} +{\rm H.c.}) &=& 
\frac{1}{2} J_{\perp} e^{-g^2} \left[ \sum_{i, j}b^{\dagger}_i  b_{j} - \sum_{i}b^{\dagger}_i  b_{i} \right] \nonumber \\
&=& \frac{1}{2}J_{\perp} e^{-g^2} \Big[ N \hat{n}_0 - 
 \hat{N_p} \Big]  \nonumber \\
&=&\sum_{k_1} \epsilon_{k_1} b^{\dagger}_{k_1} b_{k_1} , 
\end{eqnarray}
where  $ J_{\perp} = J_{\perp}^{\star}/(N-1)$,  $\hat{N_p}\equiv \sum_{k_1} b^{\dagger}_{k_1} b_{k_1}$ and 
 $ \hat{n}_0 \equiv b^{\dagger}_0 b_0$ (i.e., 
the particle number in momentum $k_1=0$ state). 
{{Thus the single particle energy is given by
\begin{eqnarray}
\epsilon_{k_1} = \frac{1}{2}J_{\perp}^{\star} \frac{N}{N-1} e^{-g^2} \delta_{{k_1},0}- \frac{1}{2}J_{\perp} e^{-g^2} .
\label{ek}
\end{eqnarray}

 We take  
 the total number of HCBs to be conserved; then,
 only the hopping term in $H^{L}_s$ will contribute to the particle excitation energy [see Eq. (\ref{Hs})].
Thus, in Eq. (\ref{E2}), the largest value of the system excitation energy in the denominator is the maximum 
single particle excitation 
energy given by 
\begin{eqnarray}
E^s_n - E^s_0 = \frac{1}{2}J_{\perp}^{\star} \frac{N}{N-1} e^{-g^2} ,
\label{En0}
\end{eqnarray}
 which is $N$ times the hopping term $(1/2)J_{\perp} e^{-g^2}$
 in $H^{L}_s$.}}
Let $\{ |q \rangle_s\}$  denote the  complete set of 
energy eigenstates (with eigenenergies $E_q^s$) of the system Hamiltonian $H^{L}_s$;
 then we can write:
 \begin{widetext}
{\begin{eqnarray}
\!\!\!\!\!\!\!\!\!\!\!\! e^{iH^{L}_s t} H^{L}_I  e^{-iH^{L}_s t} 
= \frac{1}{2} J_{\perp} e^{-g^2}\sum_{l,j>l}\sum_{q, q^{\prime}} |q \rangle_s {_s \!\langle} q| 
 e^{iH^{L}_st} \left[  \frac{1}{N}  
\sum_{{k_1}, {k_2}}b^{\dagger}_{k_1} b_{k_2} e^{i({k_1} r_l - {k_2} r_j)} \right] e^{-iH^{L}_st} |q^{\prime} \rangle_s {_s\!\langle} q^{\prime} |  
  \{\mathcal S^{{lj}^\dagger}_+ \mathcal S^{lj}_{-}-1\}
+ {\rm H. c.} ,
\end{eqnarray}}
which implies
\begin{eqnarray}
 e^{iH^{L}_s t}~ _{ph}\langle n | H^{L}_I |m\rangle_{ph} e^{-iH^{L}_s t}
 = 
 \sum_{q, q^{\prime}} |q \rangle_s {_s\!\langle} q| ~ _{ph}\langle n | H^{L}_I |m\rangle_{ph} 
 |q^{\prime} \rangle_s {_s\!\langle} q^{\prime} | e^{i(E_q^s-E_{q^{\prime}}^s)t} ,
\label{TE}
\end{eqnarray}
where {{$|E_q^s-E_{q^{\prime}}^s| \le \frac{1}{2} J_{\perp}^{\star} (\frac{N}{N-1}) e^{-g^2}$
[based on Eq. (\ref{En0})]}}. 
Substituting Eq. (\ref{TE}) in Eq. (\ref{time}), we get
{\begin{eqnarray}
&&_{ph}\langle n |
 \tilde{H}^{L}_I(t)\tilde{H}^{L}_I(t-\tau)\tilde{\rho}_s(t) \otimes R_0|n\rangle_{ph} 
\nonumber \\
&&=
 \sum_m \sum_{q, q^{\prime}, q^{\prime \prime}} \left [ |q \rangle_s {_s\!\langle} q| ~ _{ph}\langle n | H^{L}_I |m\rangle_{ph} 
|q^{\prime} \rangle_s {_s\!\langle} q^{\prime} |~ _{ph}\langle m | H^{L}_I |n\rangle_{ph}
 |q^{\prime \prime} \rangle_s {_s\!\langle} q^{\prime\prime} | 
e^{i[(E_q^s-E_{q^{\prime}}^s)t+(E_{q^{\prime}}^s-E_{q^{\prime\prime}}^s)(t-\tau)]}   \right]
 \tilde{\rho}_s(t)\frac{e^{-\beta \omega_n}}{Z} e^{i(\omega_n-\omega_m)\tau} .
\nonumber \\ 
\label{time1}
\end{eqnarray}}
Based on the above equation, at the temperature of $\rm 0~K$, we get the following:
\begin{eqnarray}
&&  \sum_n   {_{ph}}\langle n |
 \tilde{H}^{L}_I(t)\tilde{H}^{L}_I(t-\tau)\tilde{\rho}_s(t) \otimes R_0|n\rangle_{ph} 
\nonumber \\
&& =
 \sum_m \sum_{q, q^{\prime}, q^{\prime \prime}} \Bigg [ |q \rangle_s {_s\!\langle} q| ~ _{ph}\langle 0 | H^{L}_I |m\rangle_{ph} 
|q^{\prime} \rangle_s {_s\!\langle} q^{\prime} |~ _{ph}\langle m | H^{L}_I |0 \rangle_{ph}
 |q^{\prime \prime} \rangle_s {_s\!\langle} q^{\prime\prime} | 
e^{i[(E_q^s-E_{q^{\prime}}^s)t+(E_{q^{\prime}}^s-E_{q^{\prime\prime}}^s)(t-\tau)]}   \Bigg]
  \tilde{\rho}_s(t)
 e^{-i\omega_m\tau} .
\label{time1T0}
\end{eqnarray}
\end{widetext}
Since $J_{\perp}^{\star}e^{-g^2} \ll \omega $ and since 
 the maximum value of $|E^s_{q^{\prime}} - E^s_{q^{\prime \prime}}| \le J_{\perp}^{\star} e^{-g^2}$
as well as the maximum value of $|E^s_{q} - E^s_{q^{\prime}}| \le J_{\perp}^{\star} e^{-g^2}$, the following
are valid approximations: 
{\begin{eqnarray}
e^{i(t-\tau) [\omega_m + (E^s_{q^{\prime}} - E^s_{q^{\prime \prime}})]} \approx e^{i(t-\tau) \omega_m} ,
\label{approxt1}
\end{eqnarray}}
and
{\begin{eqnarray}
e^{it [-\omega_m + (E^s_{q} - E^s_{q^{\prime}})]} \approx e^{-it \omega_m} ,
\label{approxt2}
\end{eqnarray}}
where $\omega_m$ is a positive integral multiple of $\omega$. The above approximations imply that
we do not get terms producing decay.
 Then, on using the approximations given by Eqs. (\ref{approxt1}) and (\ref{approxt2}), Eq. (\ref{time1T0}) simplifies to be 
\begin{eqnarray}
&& \sum_n {_{ph}}\langle n |
 \tilde{H}^{L}_I(t)\tilde{H}^{L}_I(t-\tau)\tilde{\rho}_s(t) \otimes R_0|n\rangle_{ph} \nonumber \\
 &&=
\sum_m {_{ph}\langle 0 | H^{L}_I |m\rangle_{ph}} ~
     {_{ph}\langle m| H^{L}_I|0\rangle_{ph}} ~ \tilde{\rho}_s(t) 
 e^{-i\omega_m\tau} .
\label{time2}
\end{eqnarray}

{{Carrying out the same analysis on the remaining (i.e., third, fourth, and fifth) terms in the 
master equation, we write Eq. (\ref{mas5}) at $\rm 0~K$ temperature 
as:
\begin{widetext}
{\begin{eqnarray}
\!\!\!\!\!\!\!\!\frac{d \tilde{\rho}_s(t)}{dt} &=& -i  \sum_{n}~ _{ph}\langle n| [\tilde{H}^{L}_I(t), 
\tilde{\rho}_s(0) \otimes R_0] |n \rangle_{ph}
\nonumber \\
&& - \sum_{m} \int_0^\infty d\tau \Big[ _{ph}\langle 0 | H^{L}_I |m\rangle_{ph}~
_{ph}\langle m| H^{L}_I|0 \rangle_{ph} ~ \tilde{\rho}_s(t) 
e^{-i\omega_m\tau} 
 - ~_{ph}\langle m | H^{L}_I |0\rangle_{ph} ~ \tilde{\rho}_s(t) ~ _{ph}\langle 0|
 H^{L}_I|m\rangle_{ph} 
e^{i\omega_m \tau}
\nonumber \\ 
&&~~~~~~~~~~~~~~~~~ - ~ _{ph}\langle m | H^{L}_I |0\rangle_{ph} ~ \tilde{\rho}_s(t) ~  _{ph}\langle 0| H^{L}_I|m\rangle_{ph} 
 e^{-i\omega_m\tau} 
 +  \tilde{\rho}_s(t)~ _{ph}\langle 0 | H^{L}_I |m\rangle_{ph}~_{ph}\langle m| H^{L}_I|0\rangle_{ph} 
 e^{i\omega_m\tau}
\Big] .
\end{eqnarray}}

Next, we evaluate the first term in the above equation and show that it is zero at the
temperature of $\rm 0~ K$.
 We observe that
{\begin{eqnarray}
 Tr_R[\tilde{H}^{L}_I(t)R_0]&=& \sum_{n}~ _{ph}\langle n |\tilde{H}^{L}_I(t)R_0|n \rangle_{ph} \nonumber \\
&=& \frac{1}{2} J  e^{-g^2}\sum_{l,j\neq l} \left[ e^{iH^{L}_s t} b^{\dagger}_l b_j e^{-iH^{L}_st}~ 
_{ph}\langle 0 |  \{\mathcal S^{{lj}^\dagger}_+ \mathcal S^{lj}_{-} 
 -1\}  |0 \rangle_{ph} \right] \nonumber \\ &=& 0 .
\end{eqnarray}}
Thus, we have $  \sum_{n}~ _{ph}\langle n| [\tilde{H}^{L}_I(t), \rho_s(0) \otimes R_0] |n \rangle_{ph}=0$
and the master equation at ${\rm 0 ~K}$ temperature simplifies as:
\begin{eqnarray}
 \frac{d \tilde{\rho}_s(t)}{dt} 
&=& - \sum_n  
\Big[ \int_0^{\infty} d\tau ~ e^{- i(\omega_n -i \eta)\tau} |_{ph}\langle 0 |H^{L}_I |n \rangle_{ph}|^2 ~ \tilde{\rho}_s(t)
+ 
\int_{0}^{\infty} d\tau ~ e^{i(\omega_n+i\eta)\tau}
 ~\tilde{\rho}_s(t) ~|_{ph}\langle 0 |H^{L}_I |n \rangle_{ph}|^2  \nonumber \\
&&~~~~~~~~~ - \int_{-\infty}^{\infty} d\tau ~ e^{i\omega_n \tau} ~ _{ph}\langle n| H^{L}_I |0\rangle_{ph} 
~\tilde{\rho}_s(t)~ _{ph}\langle 0|H^{L}_I|n\rangle_{ph} 
    \Big], 
\label{21}
\end{eqnarray}
{{where $\eta \rightarrow +0$}}.
Now, we know that $\int_{-\infty}^{\infty} d\tau e^{i \omega_n \tau} \propto \delta(\omega_n)$.
Therefore, on using this relation and  the fact that ${_{ph}\langle 0| H^{L}_I |0\rangle_{ph}} =0$, the 
third term in Eq. (\ref{21}) vanishes; hence, we get
\begin{eqnarray}
\frac{d \tilde{\rho}_s(t)}{dt} =  i~\sum_n \left[
  \frac{ |_{ph}\langle 0 |H^{L}_I |n \rangle_{ph}|^2 }{\omega_n} \tilde{\rho}_s(t) - 
\tilde{\rho}_s(t) \frac{ |_{ph}\langle 0 |H^{L}_I |n \rangle_{ph}|^2 }{\omega_n} 
\right]  .
\label{difeq}
\end{eqnarray}
{{Here it should be pointed out that the above simplified form for the master equation was possible
due to the Markovian approximation made.
Based on Eq. (\ref{H_eff}),}} we identify the term $-\sum_n \left[\frac{ |_{ph}\langle 0 |H^{L}_I |n \rangle_{ph}|^2 }{\omega_n} \right]$ in
the above equation
 to be the term $H^{(2)}$  obtained in second-order perturbation which together with $H^{L}_s$ 
makes up the effective Hamiltonian (in second-order perturbation).
{{On noting that $\tilde{\rho}_s(t) = e^{iH^{L}_st} \rho_s(t) e^{-iH^{L}_s t}$ and that
$H^{(2)}$ commutes
 with $H^{L}_s$ (see section $2$), it follows from Eq. (\ref{difeq}) that
\begin{eqnarray}
 \frac{d {\rho}_s(t)}{dt} =  -i \left[H^{L}_s+H^{(2)}, \rho_s(t) \right] .
\label{commut}
\end{eqnarray}
The above Eq. (\ref{commut}) shows that the effective Hamiltonian ($H^{L}_s+H^{(2)}$) governs the unitary evolution of the reduced density matrix
$\rho_s(t)$ with ${\rho}_s(t) = e^{-i(H^{L}_s+H^{(2)})t} \rho_s(0) e^{i(H^{L}_s+H^{(2)}) t}$. }}
 Let $|n\rangle_s$ be the simultaneous eigenstate for $H^{(2)}$ and $H^{L}_s$ with
 eigenvalues $E_n^{(2)}$ and $E_n^s$, respectively. Then, from the above Eq. (\ref{commut}) we get:
\begin{eqnarray}
 {_s\langle n|\rho_s(t) | m \rangle_s}
& =& e^{-i(E_n- E_m)t}~ _s\langle n|\rho_s(0) | m \rangle_s ,
\label{sol}
\end{eqnarray}
\end{widetext}
where $E_n= E_n^s+E_n^{(2)}$.
Thus we see from the above equation that there is only a phase shift but no decoherence!
 Thus, up to second order in perturbation, the 
 assumption $J_{\perp}^{\star} e^{-g^2} \ll \omega$, the long range
of the model, and the Markov approximation 
 ($\tau_{c} \ll \tau_{s}$) together have ensured that the 
system, with a fixed  $\sum_{i} n_i$, does not decohere; furthermore, there is no change in the population ${_s\langle n|\rho_s(t) | n \rangle_s}$.}

Extending the dynamical analysis to the non-Markovian case for a system with different site 
energies is difficult for a many-site and many-body situation; hence, in the next section, we restrict ourselves
to a two-site system involving a single HCB. 

\section{Non-Markovian analysis for a two-site case with different site energies}
\subsection{Single-mode case}
Here we consider the case where one HCB is hopping between two sites each having its local phonon environment. Initially, for simplicity,
we consider the baths and the interaction terms to involve only a single mode and ignore the wavenumber dependence.
 The model Hamiltonian is given by
\begin{eqnarray}
 H&=&\varepsilon_{1} (n_{1}-\frac{1}{2})+\varepsilon_{2} (n_{2}-\frac{1}{2})-\frac{J_{\perp}}{2}(b_{1}^{\dagger}b_{2}+b_{2}^{\dagger}b_{1})\nonumber \\
&& +J_{\parallel}(n_{1}-\frac{1}{2})(n_{2}-\frac{1}{2})
 +g\omega\sum_{i=1,2} (n_{i}-\frac{1}{2})( a_i + a_{i}^{\dagger})\nonumber \\
 &&+ \omega \sum_{i=1,2} a_{i}^{\dagger} a_i,
 \end{eqnarray}
where $\varepsilon_{1}$ and $\varepsilon_{2}$ are the site energies.
 In the regime of strong electron-phonon coupling, 
we perform the LF transformation 
\begin{equation}
 H^{L}=e^{S}He^{-S},
\end{equation}
where again $S= -\sum_{i}g (n_i-\frac{1}{2}) (a_i - a_{i}^{\dagger})$.
We make our analysis in the polaronic frame of reference where the system Hamiltonian $H_{s}^{L}$, interaction 
Hamiltonian $H^{L}_{I}$ and the displaced-phonon Hamiltonian $H^{L}_B$ are given by
\begin{eqnarray}
 H_s^{L}&=&\varepsilon_{1} (n_{1}-\frac{1}{2})+\varepsilon_{2} (n_{2}-\frac{1}{2})
-  \frac{J_\perp e^{-g^2}}{2} (b_{1}^{\dagger}b_{2}+b_{2}^{\dagger}b_{1})
\nonumber \\
&& + J_\lVert (n_{1}-\frac{1}{2})(n_{2}-\frac{1}{2}) ,
\label{hsl}
\end{eqnarray}
\begin{eqnarray}
 H_I^{L}&=& -\frac{1}{2} [J_\perp ^+ b_{1}^{\dagger}b_{2} + J_\perp ^- b_{2}^{\dagger}b_{1} ] ,
\label{hil}
\end{eqnarray}
and 
\begin{eqnarray}
 H_{env}^{L} &=& \omega \sum_{i=1,2} a_{i}^{\dagger} a_i ,
\label{hbl}
\end{eqnarray}
respectively. In the above equations
 \begin{eqnarray}
 J_\perp ^{\pm} &=& J_\perp e^{\pm g[(a_2 - a_{2}^{\dagger})-(a_1 - a_{1}^{\dagger})]} -  J_\perp e^{-g^2} ,
\end{eqnarray}
and
\begin{eqnarray}
 J_\perp e^{-g^2} &=& _{ph}\langle 0 |J_\perp e^{\pm g[(a_2 - a_{2}^{\dagger})-(a_1 - a_{1}^{\dagger})]}| 0 \rangle_{ph} .
\end{eqnarray}

The system Hamiltonian $H_{s}^{L}$ represents HCB coupled to the mean-phonon field and  $\frac{1}{2}J_{\perp}e^{-g^2}$ is the resulting renormalized
hopping amplitude. 
In the subspace involving only one HCB and two sites, 
the two eigenstates of $H_s^{L}$ are given by $\frac{1}{\sqrt{1+\chi_{1}^2}}(\chi_1|10\rangle+|01\rangle)$ and 
$\frac{1}{\sqrt{1+\chi_{2}^2}}(\chi_2|10\rangle+|01\rangle)$ with 
corresponding eigenenergies $\frac{-J_{\parallel}-2\sqrt{\Delta\varepsilon^2+J_\perp ^2 e^{-2g^2}}}
{4}$ and $\frac{-J_{\parallel}+2\sqrt{\Delta\varepsilon^2+J_\perp ^2 e^{-2g^2}}}{4}$, respectively; here, 
$\chi_1=-\frac{\Delta\varepsilon-\sqrt{\Delta\varepsilon^2+
J_\perp ^2 e^{-2g^2}}}{J_\perp  e^{-g^2}}$ and $\chi_2=-\frac{\Delta\varepsilon+\sqrt{\Delta\varepsilon^2+
J_\perp ^2 e^{-2g^2}}}{J_\perp  e^{-g^2}}$ with site-energy difference
$\Delta\varepsilon=\varepsilon_1-\varepsilon_2$. 
The interaction 
Hamiltonian $H_I^{L}$ represents the HCBs coupled to fluctuations of local phonons around their mean-phonon field. 
As the interaction in the polaronic frame of reference 
is weak (compared to that in the original frame of reference), one can treat $H_I^{L}$ as a perturbation. To analyze the non-Markovian dynamics of 
the model, we start with the simply separable initial state  $\rho_T(0) = \rho_s(0) \otimes R_0$ where 
$R_0$  is the 
phonon density matrix at thermal equilibrium and is given by $R_0=\sum_{n_1,n_2}|n_1,n_2 \rangle_{ph}~\!_{ph}\langle n_1,n_2|e^{-\beta \omega_{n_1,n_2}}/Z$. 
Here, $n_1$ and $n_2$ are the phonon occupation numbers at sites 1 and 2, respectively; henceforth, we will use 
the notation $|n\rangle_{ph} \equiv |n_1,n_2 \rangle_{ph}$ and $\omega_n \equiv \omega_{n_1,n_2} = \omega(n_1+n_2)$.

\begin{figure}
\centering
\begin{overpic}[angle=90,angle=90,angle=90,width=3.2in,height=2.5in]{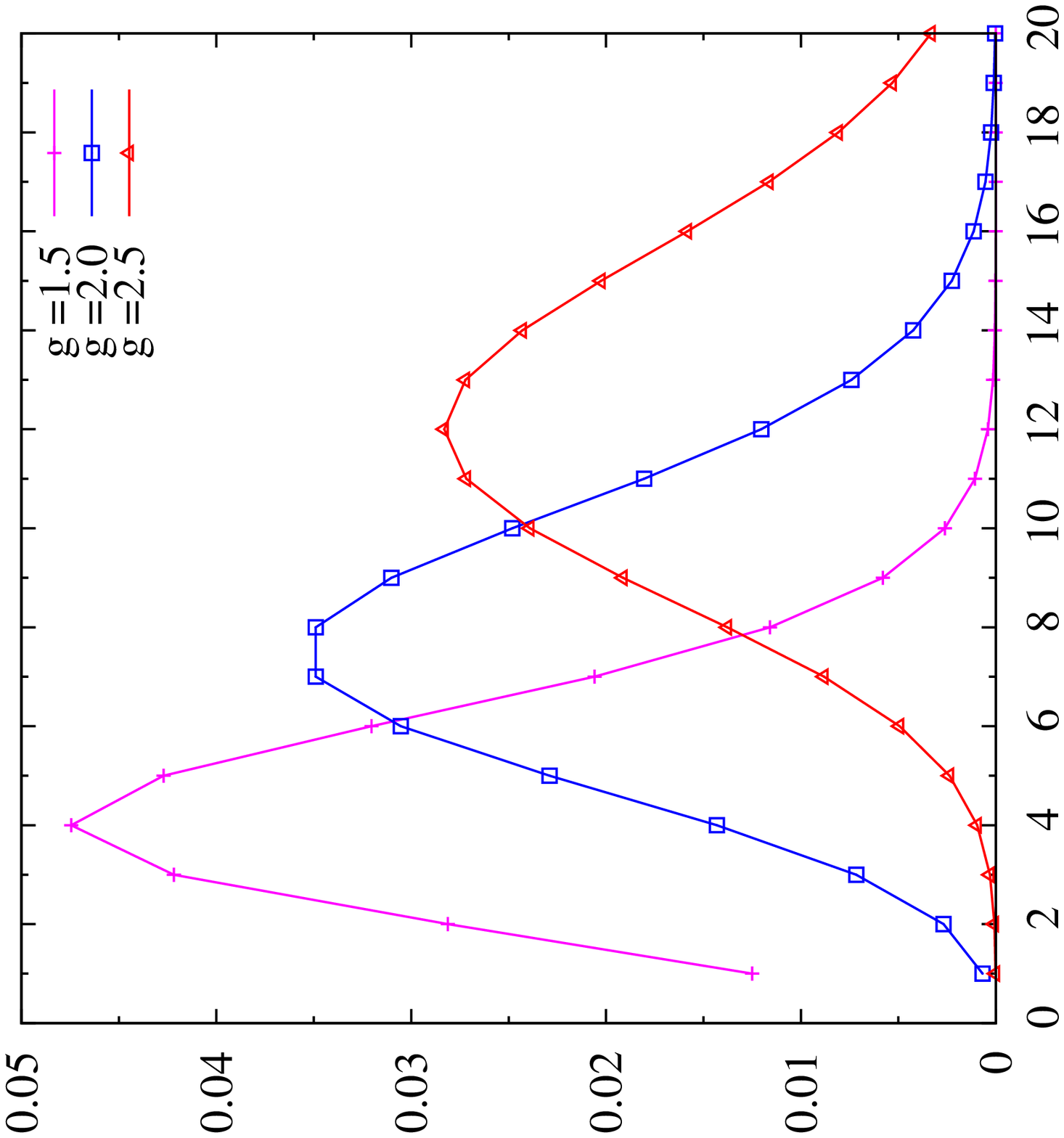}
\put(1,39){\rotatebox{90}{$ \mathbb{C^N} $}}
\put(54,-5){$\mathbb{N}$}
\end{overpic}
\caption{${\mathbb{ C^N}}$ as a function of $\mathbb{N}$ for different values of coupling
$g$
and for $\frac{J_{\perp}}{\omega}=1.0$ .}
\label{fig1}
\end{figure}

\begin{figure}
\centerline{\includegraphics[width=3.3in,height=0.9in]{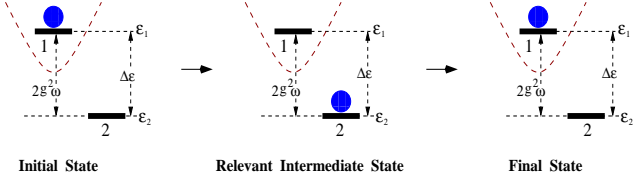}}
\caption{Schematic representation of the site energies and hopping process leading to minimum coherence ${\rm C(t)}$ and
maximum decay of excited state ${\rm P(t)}$; the intermediate state gives the dominant contribution.
Here location of the HCB is given by the filled circle.
Parabolic curve
at site 1 depicts full distortion of the lattice environment at that site with
corresponding energy $-g^2\omega$ ($+g^2\omega$ ) if the HCB is present
(absent) at that site.
}
\label{fig2}
\end{figure}

 Now, we start with the second-order, time-convolutionless (TCL), non-Markovian, quantum-master equation 
\begin{eqnarray}
\!\!\!\!\!\! {\frac{d \tilde{\rho}_s(t)}{dt} 
= - \int_0^t d\tau Tr_R[\tilde{H}_I^L(t),[\tilde{H}_I^L(\tau), \tilde{\rho}_s(t) \otimes R_0]] .}
\label{mas4}
\end{eqnarray}
At zero temperature, Eq. (\ref{mas4}) can be rewritten as 
\begin{widetext}
\begin{eqnarray}
\!\!\!\!\!\! \frac{d \tilde{\rho}_s(t)}{dt} &=& - \sum_{m} \int_0^t d\tau \Big[~ _{ph}\langle 0 |
 \tilde{H}_I^L(t)| m \rangle_{ph}~ _{ph}\langle m| \tilde{H}_I^L(\tau)| 0 \rangle_{ph} \tilde{\rho}_s(t)   
- {~_{ph}}\langle m| \tilde{H}_I^L(t)|0\rangle_{ph}\tilde{\rho}_s(t) {_{ph}}\langle 0 | \tilde{H}_I^L(\tau) |m\rangle_{ph} \nonumber \\
 && ~~~~~~~~~~~~~~~~~~~- {~_{ph}}\langle m| \tilde{H}_I^L(\tau)| 0\rangle_{ph} \tilde{\rho}_s(t) {_{ph}}\langle 0| \tilde{H}_I^L(t) |m\rangle_{ph} 
 + 
~ \tilde{\rho}_s(t)  _{ph}\langle 0|
 \tilde{H}_I^L(\tau)| m \rangle_{ph}~ _{ph}\langle m | \tilde{H}_I^L(t)| 0\rangle_{ph}  
 \Big] .
\label{mas}
\end{eqnarray}
\end{widetext}
We choose the basis $\{|10\rangle,|01\rangle\}$ for our analysis and obtain the following
useful expressions: 
\begin{eqnarray}
 e^{-i H_s^{L}t}|10\rangle&=&[p(t)^* |10\rangle -i\kappa q(t) |01\rangle]e^{i\frac{J_{\parallel}}{4}t} ,
\label{basis1}
\end{eqnarray}
and
\begin{eqnarray}
 e^{-i H_s^{L}t}|01\rangle&=&[p(t) |01\rangle -i\kappa q(t) |10\rangle]e^{i\frac{J_{\parallel}}{4}t},
 \label{basis2}
\end{eqnarray}
where $p(t)=\rm {cos}\Big (t \sqrt{\frac{\Delta \varepsilon^2}{4}+\kappa^2} \Big)+i\frac{\Delta \varepsilon}{2}
 \frac{\rm{sin}\Big ( t \sqrt{\frac{\Delta \varepsilon^2}{4}+\kappa^2} \Big)}{\sqrt{\frac{\Delta \varepsilon^2}{4}+\kappa^2}}$,
 $q(t)=\frac{\rm{sin}\Big (t \sqrt{\frac{\Delta \varepsilon^2}{4}+\kappa^2} \Big)}{\sqrt{ \frac{\Delta \varepsilon^2}{4}+\kappa^2}}$ 
 and $\kappa=-\frac{J_{\perp} e^{-g^2}}{2}$. In addition, we also evaluate the matrix element 
 \begin{eqnarray}
   _{ph}\langle 0 |H^{L}_{I}| n \rangle_{ph}&=& \kappa  (-1)^{n_1} \sqrt{C_{n}}
\Big (b_{1}^{\dagger}b_{2}+(-1)^{n_1+n_2} b_{2}^{\dagger}b_{1} \Big), 
  \nonumber \\
  \label{matelement}
 \end{eqnarray}
where $C_n=\frac{g^{2(n_1 + n_2)}}{{n_1 !  n_2 !}}$.
Taking the matrix elements, with respect
 to $|10\rangle$ and $|01\rangle$, on both sides of Eq. (\ref{mas}) and
 by using Eqs. (\ref{basis1})--(\ref{matelement}),
 we calculate the matrix elements of the four terms on the right hand side of Eq. (\ref{mas}) (details
are shown in the Appendix). 
 For the case when $|\Delta \varepsilon| \gg |\kappa|$, $|10\rangle$ and $|01\rangle$ 
are the approximate eigenstates of $H^{L}_s$. In this regime of parameter 
 values,  we ignore the ratio $\frac{|\kappa|}{|\Delta \varepsilon|}$ compared to 1 
[in Eqs. (\ref{1st})--(\ref{4th})] and finally
 obtain simple form of the master equation for $\langle 10 | \tilde{\rho}_s(t)|01\rangle$: 
 \begin{widetext}
 \begin{eqnarray}
  \frac{d \langle 10 | \tilde{\rho}_s(t)|01\rangle }{dt}&=&-i\kappa^2 \sum^{\prime}_{n} C_{n} \Big[\langle 10 | \tilde{\rho}_s(t)|01\rangle
  \Big(\frac{e^{-i(\omega_n -\Delta\varepsilon)t}}{\omega_n -\Delta\varepsilon} -\frac{e^{i(\omega_n +\Delta\varepsilon)t}}{\omega_n+\Delta\varepsilon}
  -\frac{2\Delta\varepsilon}{\omega^{2}_n-\Delta\varepsilon^2}\Big) \nonumber \\ 
 &&~~~~~~~~~~~~~~~~~ + \langle 01 | \tilde{\rho}_s(t)|10\rangle (-1)^{n_1+n_2}e^{2i\Delta \varepsilon t} \Big( \frac{e^{i(\omega_n -\Delta\varepsilon)t}}{\omega_n -
  \Delta\varepsilon} -\frac{e^{-i(\omega_n +\Delta\varepsilon)t}}{\omega_n+\Delta\varepsilon}
  -\frac{2\Delta\varepsilon}{\omega^{2}_n-\Delta\varepsilon^2}    \Big)  \Big] ,
  \label{offeqn}
  \end{eqnarray}
  and its complex conjugate equation for $\langle 01 | \tilde{\rho}_s(t)|10\rangle$. 
In the above equation, $\sum^{\prime}_{n}\equiv \sum^{\prime}_{n_1,n_2}$ excludes the case where $n_1$ and $n_2$ 
  are simultaneously zero.
  Similarly, for the diagonal element $\langle 10 | \tilde{\rho}_s(t)|10\rangle$, 
  the differential equation can be written as
  \begin{eqnarray}
    \frac{d \langle 10 | \tilde{\rho}_s(t)|10\rangle }{dt}&=&-2\kappa^2\sum^{\prime}_{n} C_{n} \Big[ \langle 10 | \tilde{\rho}_s(t)|10\rangle
    \Big(\frac{{\rm {sin}}(\omega_n +\Delta\varepsilon)t}{\omega_n +\Delta\varepsilon}+\frac{{\rm {sin}}(\omega_n -\Delta\varepsilon)t}
    {\omega_n -\Delta\varepsilon} \Big)-\frac{{\rm {sin}}(\omega_n +\Delta\varepsilon)t}{\omega_n +\Delta\varepsilon} \Big].
    \label{diageqn}
  \end{eqnarray}
 The above Eq. (\ref{diageqn}) is a first-order, non-homogeneous, differential equation; its 
 solution is given by
\begin{eqnarray}
 \!\!\!\! \langle 10 |{\rho}_s(t)|10\rangle&=&\langle 10 |{\rho}_s(0)|10\rangle {\rm exp}\Big[-2\kappa^2\sum^{\prime}_{n}C_n \Big( \frac{1-{\rm cos}(\omega_n +\Delta\varepsilon)t}
  {(\omega_n +\Delta\varepsilon)^2}+\frac{1-{\rm cos}(\omega_n -\Delta\varepsilon)t}
  {(\omega_n -\Delta\varepsilon)^2}\Big)\Big] \nonumber \\
  &&+2\kappa^2{\rm exp}\Big[-2\kappa^2\sum^{\prime}_{n}C_n \Big( \frac{1-{\rm cos}(\omega_n +\Delta\varepsilon)t}
  {(\omega_n +\Delta\varepsilon)^2}+\frac{1-{\rm cos}(\omega_n -\Delta\varepsilon)t}
  {(\omega_n -\Delta\varepsilon)^2}\Big)\Big] \nonumber \\
 &&~ \times\sum^{\prime}_{n} C_n \int^t_{0} dt^{\prime}~\frac{{\rm {sin}}(\omega_n +\Delta\varepsilon)t^{\prime}}{\omega_n +\Delta\varepsilon}
  {\rm exp}\Big[2\kappa^2\sum^{\prime}_{n}C_n \Big( \frac{1-{\rm cos}(\omega_n +\Delta\varepsilon)t^{\prime}}
  {(\omega_n +\Delta\varepsilon)^2}+\frac{1-{\rm cos}(\omega_n -\Delta\varepsilon)t^{\prime}}
  {(\omega_n -\Delta\varepsilon)^2}\Big)\Big] .
  \label{diasoln}
\end{eqnarray}
\end{widetext}
The solution Eq. (\ref{diasoln}) has a part dependent on the initial value of $\langle 10 |{\rho}_s(t)|10\rangle$ and 
a part independent of that.
 
To understand decoherence  and the decay of the excited state ($|10\rangle$),  we define two quantities: the coherence factor 
${\rm C(t)}=\frac{ |\langle 10 |{\rho}_s(t)|01\rangle|}{ |\langle 10 |{\rho}_s(0)|01\rangle|}$ 
and the population of the excited state  ${\rm P(t)}= \langle 10 |{\rho}_s(t)|10\rangle$.
We numerically solve the coupled differential equations given by Eq. (\ref{offeqn}) and its complex-conjugate equation and 
plot the dynamical behavior of ${\rm C(t) }$ in Figs. \ref{fig3}, \ref{fig5}(a), and \ref{fig7}.
We also depict the time dependence of ${\rm P(t)}$ in Figs. \ref{fig4}, \ref{fig5}(b), and \ref{fig8}.
We analyze below Figs. \ref{fig3}--\ref{fig8} and show that the period of oscillation and the amplitude 
of oscillation of both ${\rm C(t)}$ and ${\rm P(t)}$ increase as the site energy difference $  \Delta\varepsilon$ approaches a harmonic $\omega_n$;
also, the closer the $\Delta\varepsilon$ is to $\omega_n$, the smaller are the equilibrium values of ${\rm C(t)}$ and ${\rm P(t)}$. 
Furthermore, the closer $\omega_n$ is to $2g^2\omega$ (i.e., twice the polaronic energy), the more prominent
are the period and amplitude of oscillations.  Interestingly too, we find that the stronger the coupling $g$, the weaker is
the decoherence and the decay of the excited state population.

\begin{figure}
\centerline{\includegraphics[angle=90,angle=90,angle=90,width=2.6in,height=3.5in]{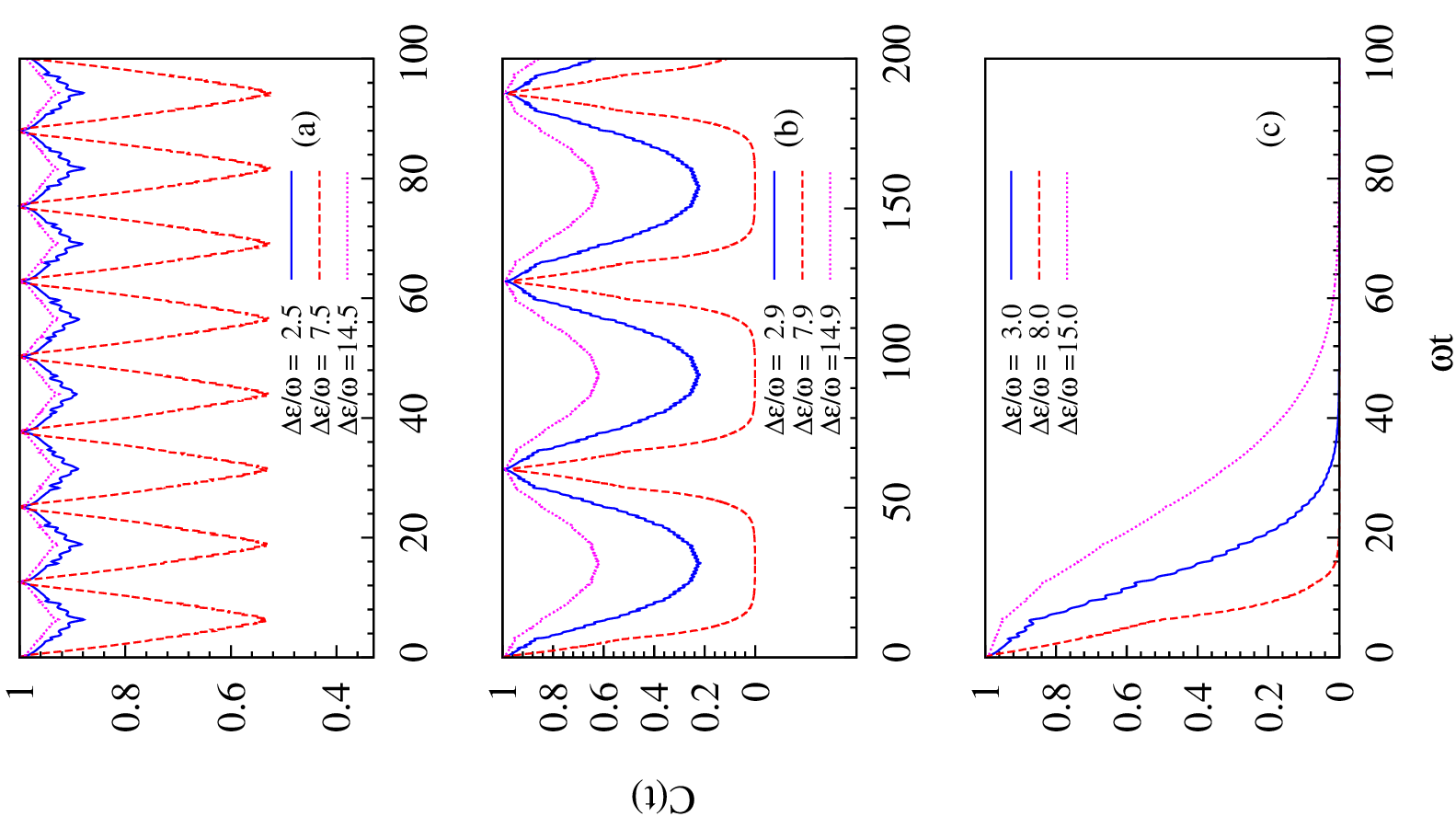}}
\caption{Time dependence of ${\rm C(t)}$ for $\frac{J_{\perp}}{\omega}=1.0$, $g=2.0$, and when
 (a) $\frac{\Delta \varepsilon}{\omega}=2.5, 7.5$ and $14.5$; 
(b) $\frac{\Delta \varepsilon}{\omega}=2.9, 7.9$ and $14.9$; and (c) $\frac{\Delta \varepsilon}{\omega}=3.0, 8.0$ and $15.0$ .}
\label{fig3}
\end{figure}

\begin{figure}
\centerline{\includegraphics[angle=90,angle=90,angle=90,width=2.6in,height=3.5in]{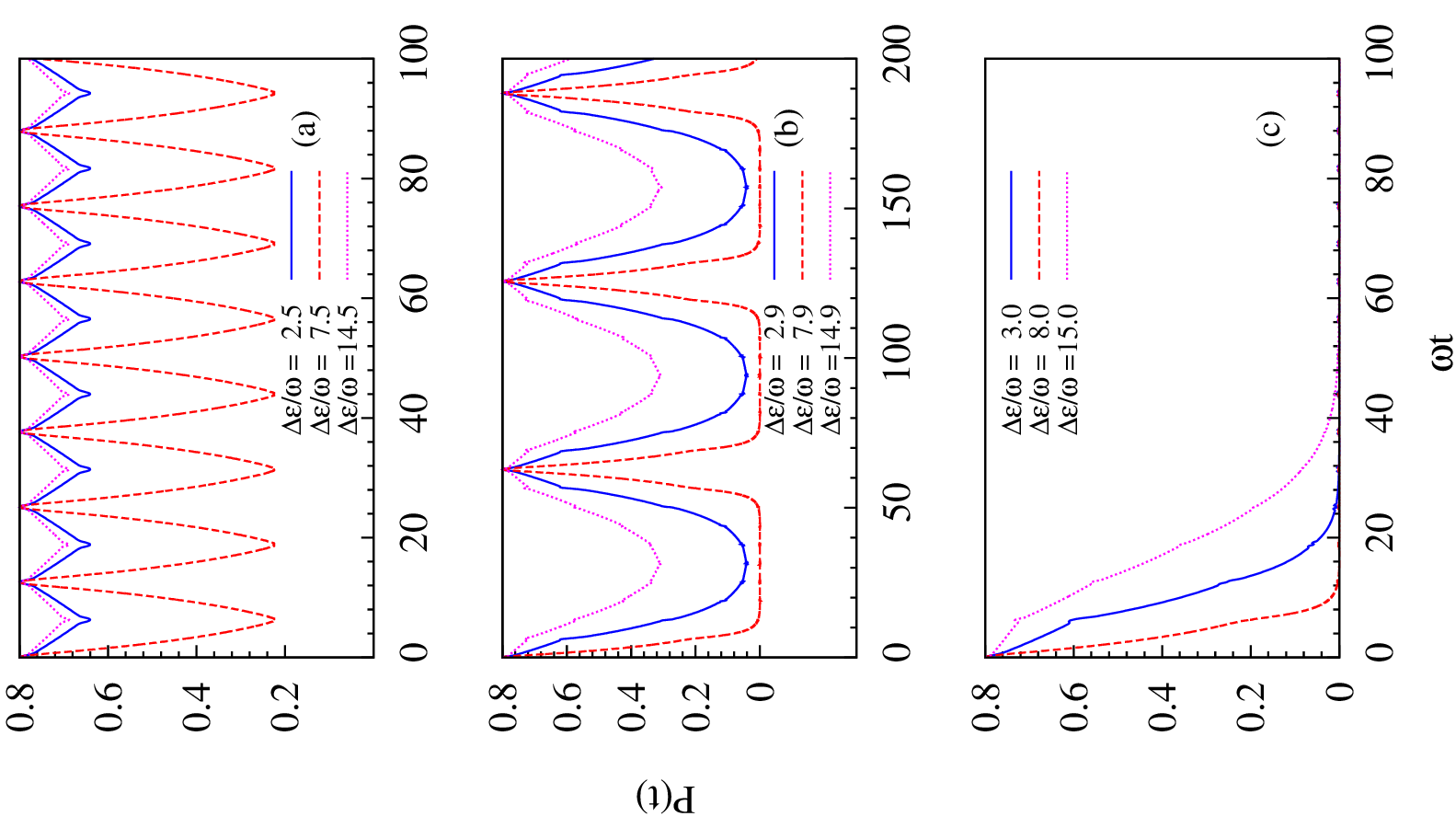}}
\caption{Time dependence of ${\rm P(t)}$ for  $\frac{J_{\perp}}{\omega}=1.0$, ${\rm P(0)}=0.8$, $g=2.0$, and when
(a) $\frac{\Delta \varepsilon}{\omega}=2.5, 7.5$ and $14.5$; 
(b) $\frac{\Delta \varepsilon}{\omega}=2.9, 7.9$ and $14.9$; and
(c) $\frac{\Delta \varepsilon}{\omega}=3.0, 8.0$ and $15.0$.}
\label{fig4}
\end{figure}
In Figs. \ref{fig3} and \ref{fig4}, we study three cases of proximity of 
$\Delta\varepsilon$ to $\omega_n$: $\frac{\Delta \varepsilon}{\omega}=2.5,~7.5,~ \&~14.5$; 
$\frac{\Delta \varepsilon}{\omega}=2.9,~7.9,~ \&~14.9$; 
and $\frac{\Delta \varepsilon}{\omega}= 3.0,~8.0,~ \&~15.0$.
One can see from Eq. (\ref{offeqn}) that, for values of $\Delta \varepsilon$ close to ${\omega_n}$ (i.e., 
for $\frac{\Delta \varepsilon}{\omega}=2.9,~7.9,~\&~14.9$), 
the dominant terms have arguments of the periodic functions being given by   
$(\omega_n -\Delta\varepsilon)t=0.1\omega t$; consequently, there is a large period of oscillation
(given by $\omega t = 20\pi$) in Fig. \ref{fig3}(b). Also,
 the amplitude of oscillation is dominated by the term $\sin[(\omega_n -\Delta\varepsilon)t]/(\omega_n -\Delta\varepsilon)$ 
and hence the amplitude increases with decreasing values of $(\omega_n -\Delta\varepsilon)$. 
{Furthermore, the coherence factor also depends on
the number of degenerate phonon states with eigenenergy $\omega_n$; 
the contribution of this degeneracy [as seen from Eq. (\ref{offeqn})]
 is determined by the term 
\begin{eqnarray}
\mathbb{C^ N}=\frac{\kappa^2}{\omega^2}\sum_{n_1,n_2; (n_1+n_2)={\mathbb{N}}} C_n=\frac{\kappa^2}{\omega^2}\frac{{(2g^2)}^{{\mathbb {N}}}}{{\mathbb {N}}!},
\end{eqnarray}
where $\sum_{n_1,n_2;(n_1+n_2)={\mathbb{N}}}$ adds up all $C_n$ with $(n_1+n_2)={\mathbb{N}}$. 
The closeness of $\Delta \varepsilon$ to $\omega_n$ and the value of $\mathbb{C^{N}}$ together determine the strength of decoherence. 
The value of  $\mathbb{C^N}$ increases with increasing ${\mathbb{N}}=n_1+n_2$ up to some limit as depicted in Fig. \ref{fig1}.
One can also see that the maximum of $\mathbb{C^N}$ occurs at phonon eigenenergy 
$\omega_n$  close to $2g^2\omega$. In other words, for a particular $g$, 
when 
 $\Delta\varepsilon$ is close to twice the polaronic 
energy $g^2\omega$, decoherence is maximum with $\omega_n$ closest to $2g^2\omega$ making the dominant contribution.} 
{In second-order perturbation picture, the relevant process involves the particle hopping 
from one site to another and coming back to the 
initial site. Now, the initial state is described by the occupied site with polaronic energy 
(lattice distortion potential) $-g^2 \omega$; whereas the 
intermediate state for perturbation theory corresponds to the occupied second-site being
 without distortion and the unoccupied first-site having distortion energy $+g^2 \omega$ 
[see Fig. \ref{fig2} given above and Fig. 2(a) in Ref. \onlinecite{srsypbl}]. 
As the energy difference between the initial and the intermediate states
[i.e.,  $(\varepsilon_1-g^2\omega)-(\varepsilon_2+g^2\omega)=\Delta\varepsilon-2g^2\omega$] 
approaches zero, the hopping process becomes more dominant leading to stronger decoherence.}
The above observations that the period of oscillation being inversely proportional
to $\Delta \varepsilon - \omega_n$ and that the values of  $1/(\Delta \varepsilon - \omega_n)$  and $\mathbb{C^{N}}$ together 
determine the strength of decoherence 
are also exemplified for the cases when $\Delta\varepsilon =\omega_n$ [through  Fig. \ref{fig3}(c)
when $\frac{\Delta \varepsilon}{\omega}= 3.0,~8.0,~ \&~15.0$] and when $|\Delta\varepsilon -\omega_n| =\omega/2$ [through
Fig. \ref{fig3}(a) when 
$\frac{\Delta \varepsilon}{\omega}=2.5,~7.5,~ \&~14.5$].
 In Fig. \ref{fig3}(c) [\ref{fig3}(a)],
the period of oscillation is infinity [$4\pi/\omega$] and the decoherence is stronger [weaker] than in Fig. \ref{fig3}(b).
{It should be clear that recoherence occurs in Fig. \ref{fig3} because we are dealing with single mode phonons; the closer
that $\Delta \varepsilon$ 
approaches $\omega_n$, the later does the recoherence occur (i.e., recoherence time is inversely proportional to 
$\Delta \varepsilon  - \omega_n$.).
}

\begin{figure}
\centerline{\includegraphics[angle=90,angle=90,angle=90,width=3.2in,height=2.5in]{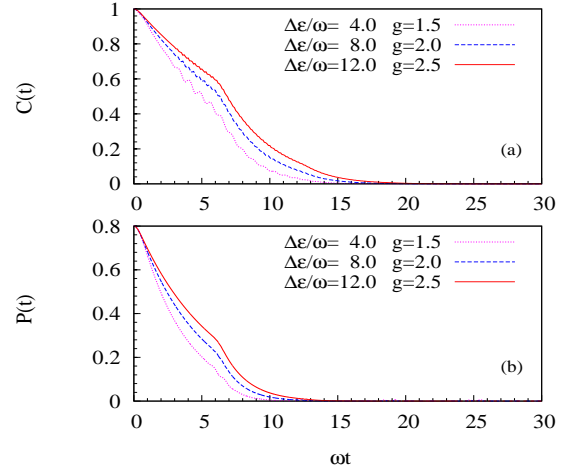}}
\caption{Time dependence of (a) {\rm C(t)} and (b)  {\rm P(t)} [with ${\rm P(0)}=0.8$] for different values of coupling $g$
and $\frac{J_{\perp}}{\omega}=1.0$ when $\Delta \varepsilon/\omega$ takes integer values closest to $2g^2$.}
\label{fig5}
\end{figure}

Similar to the above analysis of Eq. (\ref{offeqn}), one can analyze  Eq. (\ref{diasoln}) to gain an understanding
of ${\rm P(t)}$. 
For comparatively large initial values  $\langle 10 |{\rho}_s(0)|10\rangle$, 
the time dependence of $ \langle 10 |{\rho}_s(t)|10\rangle$ is mainly determined by the homogeneous part in Eq. (\ref{diasoln}).
{The role of the inhomogeneous term
 can be understood from Fig. \ref{fig6} by taking  ${\rm P(0)}=0$ in Eq. (\ref{diasoln}). One can see a very small variation 
of the excited state population and the peak values of oscillations in Fig. \ref{fig6}
are comparable to the order of $\frac{|\kappa|}{|\Delta\varepsilon|}$ [i.e., $\sim {\textit O}(10^{-2})$];  
we have neglected $\frac{|\kappa|}{|\Delta\varepsilon|}$ compared to 1 in our calculations. So [when ${\rm P(0)}=0$], 
 we can say that the system stays in the ground state for all 
practical purposes.
Next, in Eq. (\ref{diasoln}), we see that the homogeneous part is dominated by the oscillatory terms  
with period of oscillation being inversely proportional
to $\Delta \varepsilon - \omega_n$; here too  the values of  $1/(\Delta \varepsilon - \omega_n)$  and $\mathbb{C^{N}}$ together 
determine the strength of decay of ${\rm P(t)}$ as can be seen from Figs. \ref{fig4} (a), (b), and (c).}

To understand the dependence of ${\rm C(t)}$ and ${\rm P(t)}$ on the strength of coupling,
we study the variation of $\mathbb{C^{N}}$ on $g$ in Fig. \ref{fig1}.
The peak value of $\mathbb{C^{ N}}$ decreases with increasing $g$, i.e., the maximum decoherence/decay (which occurs when $\Delta\varepsilon = 2g^2\omega$)
 decreases as the coupling becomes stronger.
In Figs. \ref{fig5}(a) and (b),
 respectively, ${\rm C(t)}$ and ${\rm P(t)}$ are plotted for different values of $g$  with $\frac{\Delta \varepsilon}{\omega}$ taking integer values closest to 
$2g^2$.}}

\begin{figure}
\centerline{\includegraphics[angle=90,angle=90,angle=90,width=3.2in,height=1.5in]{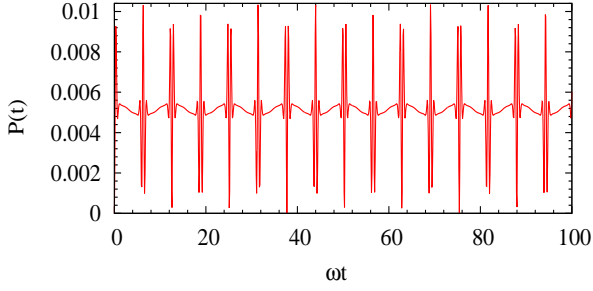}}
\caption{Time dependence of ${\rm P(t)}$ for $\frac{\Delta \varepsilon}{\omega}=2.5$ when $\frac{J_{\perp}}{\omega}=1.0$, ${\rm P(0)}=0$, and $g=2.0$.}
\label{fig6}
\end{figure}

 Now, we like to determine the values of ${\rm C(t)}$ and ${\rm P(t)}$ at  
long times, i.e., at times much larger than the largest timescale in the process $\hbar/J_{\perp} e^{-g^2}$. 
To this end, in Eqs. (\ref{offeqn}) and (\ref{diageqn}), we multiply 
 the oscillatory terms 
 with a decay term $e^{-\eta t}$
 (where $\eta\rightarrow0^{+}$) 
and rewrite the equations as
\begin{widetext}
\begin{eqnarray}
  \frac{d \langle 10 | \tilde{\rho}_s(t)|01\rangle }{dt}&=&-i\kappa^2 \sum^{\prime}_{n} C_{n} \Big[\langle 10 | \tilde{\rho}_s(t)|01\rangle
  \Big(\frac{e^{-i(\omega_n -\Delta\varepsilon-i\eta)t}}{\omega_n -\Delta\varepsilon} -\frac{e^{i(\omega_n +\Delta\varepsilon+i\eta)t}}{\omega_n+\Delta\varepsilon}
  -\frac{2\Delta\varepsilon}{\omega^{2}_n-\Delta\varepsilon^2}\Big) \nonumber \\ 
 &&~~~~~~~~~~~~~~~~~ + \langle 01 | \tilde{\rho}_s(t)|10\rangle (-1)^{n_1+n_2}e^{i(2\Delta \varepsilon+i\eta) t} \Big( \frac{e^{i(\omega_n -\Delta\varepsilon)t}}{\omega_n -
  \Delta\varepsilon} -\frac{e^{-i(\omega_n +\Delta\varepsilon)t}}{\omega_n+\Delta\varepsilon}
  -\frac{2\Delta\varepsilon}{\omega^{2}_n-\Delta\varepsilon^2}    \Big)  \Big] ,
  \label{offeta}
\end{eqnarray}
 and
 \begin{eqnarray}
    \frac{d \langle 10 | \tilde{\rho}_s(t)|10\rangle }{dt}&=&-2\kappa^2\sum^{\prime}_{n} C_{n} e^{-\eta t}\Big[ \langle 10 | \tilde{\rho}_s(t)|10\rangle
    \Big(\frac{{\rm {sin}}(\omega_n +\Delta\varepsilon)t}{\omega_n +\Delta\varepsilon}+\frac{{\rm {sin}}(\omega_n -\Delta\varepsilon)t}
    {\omega_n -\Delta\varepsilon} \Big)-\frac{{\rm {sin}}(\omega_n +\Delta\varepsilon)t}{\omega_n +\Delta\varepsilon} \Big].
    \label{diaeta}
  \end{eqnarray} 
\end{widetext}
We plot ${\rm C(t)}$ and ${\rm P(t)}$  in Figs. \ref{fig7} and \ref{fig8}, respectively, for values of $\eta/\omega=0.01$ and $0.02$. For both the values of $\eta$,
 ${\rm C(t)}$ [as well as ${\rm P(t)}$] attain the same equilibrium value. Here we should mention that, (for the chosen values of $\eta/\omega =0.01$ and $0.02$)
although the decay term $e^{-\eta t}$ does diminish  over the period of oscillation of ${\rm C(t)}$ and ${\rm P(t)}$,
 we got the same equilibrium values  for much smaller values of $\eta$ as well.}
\begin{figure}[b]
\centerline{\includegraphics[angle=90,angle=90,angle=90,width=3.2in,height=2.5in]{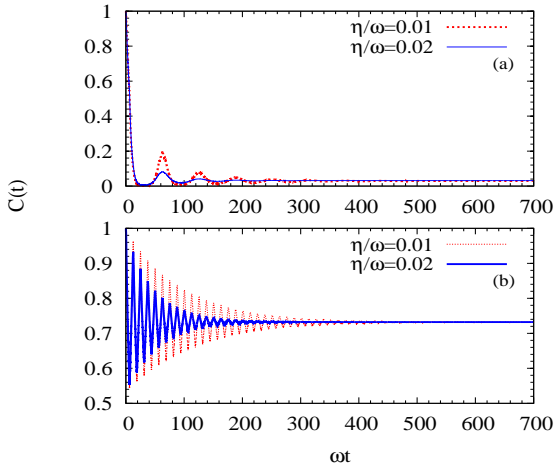}}
\caption{Time dependence of ${\rm C(t)}$ for $\frac{J_{\perp}}{\omega}=1.0$, $g=2.0$, and when
(a) $\frac{\Delta \varepsilon}{\omega}=7.9$ and (b) $\frac{\Delta \varepsilon}{\omega}=7.5$.} 
\label{fig7}
\end{figure}  
\begin{figure}[b]
\centerline{\includegraphics[angle=90,angle=90,angle=90,width=3.2in,height=2.5in]{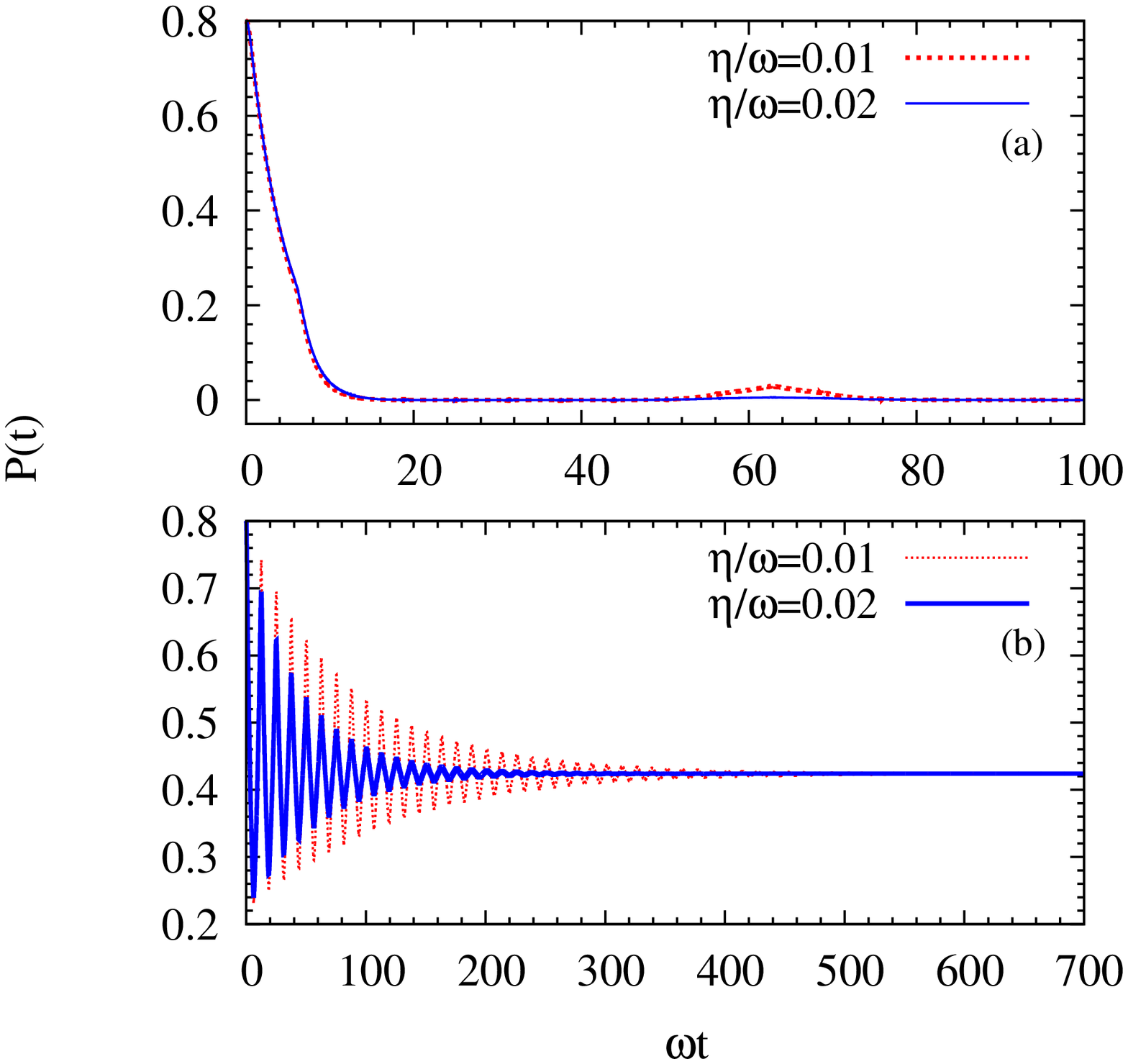}}
\caption{Time dependence of ${\rm P(t)}$ for $\frac{J_{\perp}}{\omega}=1.0$, ${\rm P(0)}=0.8$, $g=2.0$, and when
(a) $\frac{\Delta \varepsilon}{\omega}=7.9$ and (b) $\frac{\Delta \varepsilon}{\omega}=7.5$.} 
\label{fig8}
\end{figure}
  
{Lastly, we would like to compare the case of nonzero $\Delta \varepsilon$  with the case of $\Delta \varepsilon=0$  
using the plots in Figs. \ref{fig9} and \ref{fig10}. 
To analyze decoherence using ${\rm C(t)}$ and decay of ${\rm P(t)}$, for each case, we use the corresponding eigenstate basis,
 i.e., \{$|10\rangle,|01\rangle$\} for $\Delta \varepsilon\neq0$  
and \Big \{$\frac{|10\rangle-|01\rangle}{\sqrt{2}},\frac{|10\rangle+|01\rangle}{\sqrt{2}}$\Big \} for $\Delta \varepsilon=0$;
we note that $\frac{|10\rangle-|01\rangle}{\sqrt{2}}$ is the excited state for the case of $\Delta \varepsilon=0$.
We see that the periodicity of the cases with nonzero site energy [depicted in Figs. \ref{fig3}, \ref{fig4}, \ref{fig9}(a) and \ref{fig10}(a)]
is determined by the closeness of $\Delta \varepsilon$ to $\omega_n$ whereas the periodicity of the case with $\Delta \varepsilon=0$ 
 is determined by $\omega$. 
For the case of $\Delta \varepsilon=0$, in the strong coupling regime, since the system excitation energy $J_{\perp}e^{-g^2}$
is much smaller than $\omega$,
there is no possibility of energy exchange between the system and phonons. 
This results in smaller 
 equilibrium values of ${\rm C(t)}$ and ${\rm P(t)}$ for the case with finite $\Delta \varepsilon$  
compared to the case with $\Delta \varepsilon=0$ as shown in Table \ref{table}.
Also,  the oscillations of ${\rm C(t)}$ and ${\rm P(t)}$ are smaller for  the case $\Delta \varepsilon=0$
compared to the case of finite $\Delta \varepsilon$ 
as can be seen by comparing Figs. \ref{fig9}(b) and \ref{fig10}(b) with Figs. \ref{fig9}(a) and \ref{fig10}(a); here we chose 
$\Delta \varepsilon/\omega=2.5$ so that $\Delta \varepsilon$ is far away from the nearest eigenenergies $\omega_n=2\omega ~\&~3\omega$.
 Furthermore, in Fig. \ref{fig9}(b) [Fig.\ref{fig10}(b)] the $\omega t$ regions between two consecutive integer multiples of $2\pi$  
[$\pi$]  
 become flatter as the coupling $g$ increases; as $g$ increases, more number of excited states for phonons (with energy $\omega_n$)
 contribute and produce destructive interference of phases resulting in the flat region (see Ref. \onlinecite{polarondyn} for details) \cite{trugman}. On the other hand,
 in Figs. \ref{fig9}(a) and \ref{fig10}(a), only those states with $\omega_n$ close to $\Delta \varepsilon$ have a dominant contribution.

\begin{table}[t]
\begin{center}
{
\begin{tabular}{|c|c|c|c|c|c|}
\hline
\multicolumn {3}{|c|}{$\Delta \varepsilon/\omega=0$} &
\multicolumn {3}{|c|}{$\Delta \varepsilon/\omega=2.5$} \\ \cline{1-6} 
{ $g$} & { ${\rm C(t\rightarrow \infty)}$} & {${\rm P(t\rightarrow \infty)}$} &{ $g$} & { ${\rm C(t\rightarrow \infty)}$} & {${\rm P(t\rightarrow \infty)}$}\\
\hline
$1.5$ & $0.970$ & $0.794$ & $1.5$ & $0.916$ & $0.683$\\
\hline
$2.0$ & $0.993$  & $0.799$ & $2.0$ & $0.984$ & $0.778$  \\
\hline
\end{tabular}}
\caption{{Equilibrium values of ${\rm C(t)}$ and ${\rm P(t)}$ [with ${\rm P(0)}=0.8$] for $J_{\perp}/\omega=0.5$ and various values of $g$ when $\Delta \varepsilon/\omega=0 ~\&~2.5$.
}}
\label{table}
\end{center}
\end{table}

\begin{figure}[t]
\centerline{\includegraphics[angle=90,angle=90,angle=90,width=3.2in,height=2.5in]{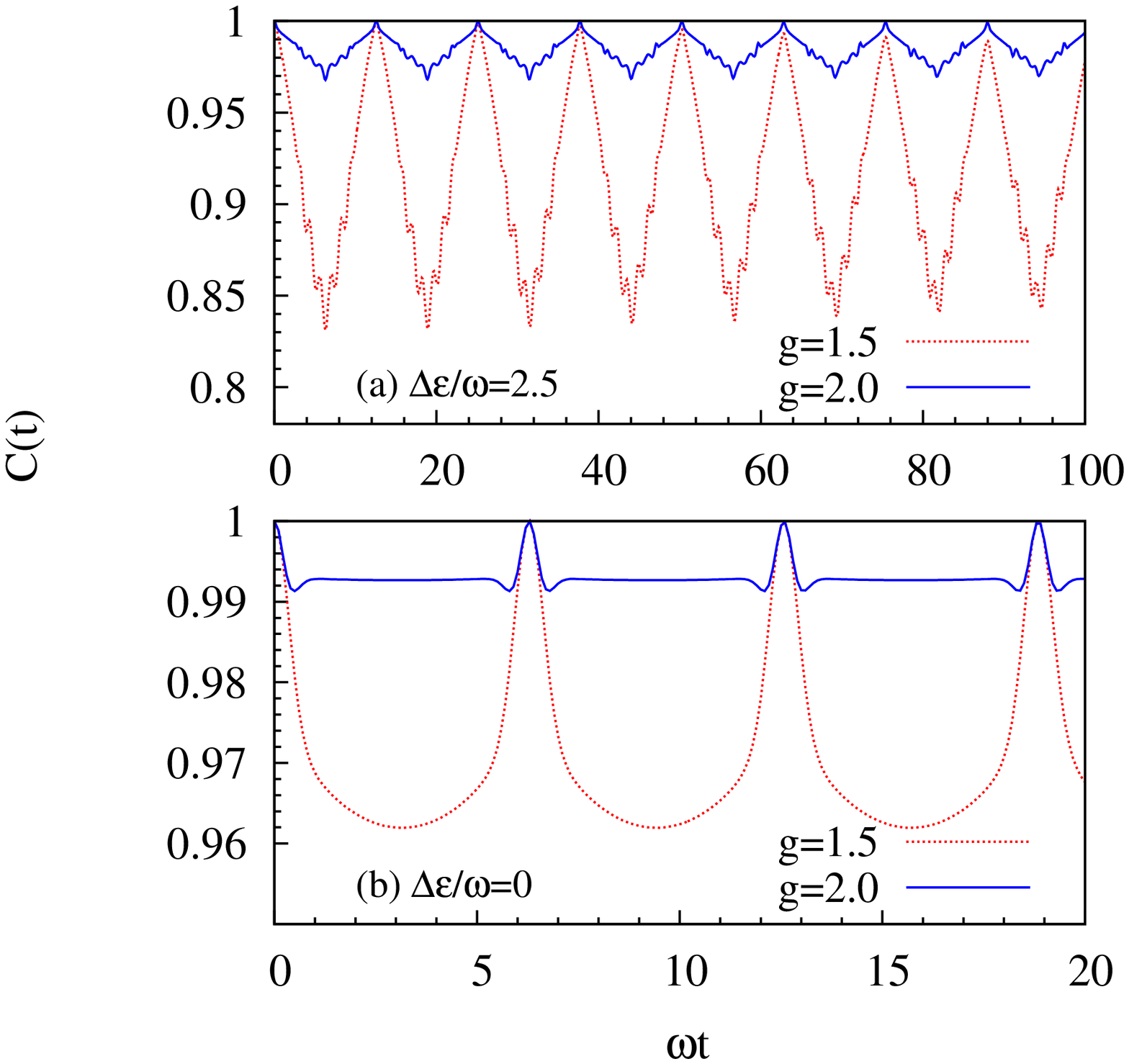}}
\caption{Time dependence of ${\rm C(t)}$ for $\frac{J_{\perp}}{\omega}=0.5$,  $g=1.5~\&~ 2.0$,
and when (a) $\frac{\Delta \varepsilon}{\omega}=2.5$ and (b) $\frac{\Delta \varepsilon}{\omega}=0$.} 
\label{fig9}
\end{figure}  

\begin{figure}[b]
\centerline{\includegraphics[angle=90,angle=90,angle=90,width=3.2in,height=2.5in]{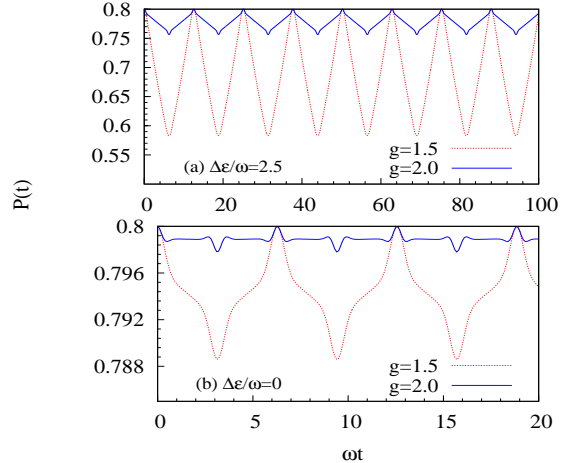}}
\caption{Time dependence of ${\rm P(t)}$ for $\frac{J_{\perp}}{\omega}=0.5$, ${\rm P(0)=0.8}$,  $g=1.5 ~\&~ 2.0$,
and when (a) $\frac{\Delta \varepsilon}{\omega}=2.5$ and (b) $\frac{\Delta \varepsilon}{\omega}=0$.} 
\label{fig10}
\end{figure}  

\subsection{Multimode case}
Here we deal with a more realistic case, i.e., we consider a continuous distribution of phonon frequencies and, for simplicity,  allow a small window 
characterized by upper and lower limits. The generalized Hamiltonian for multimode phonons in the polaronic frame of reference is written as 
\begin{eqnarray}
  H&=&\varepsilon_{1} (n_{1}-\frac{1}{2})+\varepsilon_{2} (n_{2}-\frac{1}{2})
-\frac{J_{\perp}}{2}(b_{1}^{\dagger}b_{2}+b_{2}^{\dagger}b_{1})
\nonumber \\ &&
 +J_{\parallel}(n_{1}-\frac{1}{2})(n_{2}-\frac{1}{2}) 
+ \sum_{i,k} \omega_{k} a_{i,k}^{\dagger} a_{i,k}
\nonumber \\ &&
 +\frac{1}{\sqrt{N}}\sum_{i,k}g_{k}\omega_{k}(n_{i}-\frac{1}{2})( a_{i,k} + a_{i,k}^{\dagger}) ,
 \label{multiham}
\end{eqnarray}
where $a_{i,k}$ destroys a phonon with momentum $k$ at  site $i$ and $N$ is the number of phonon modes.
To perform perturbation theory with ease, we perform Lang-Firsov transformation $H^{L}=e^{S}He^{-S}=H_s^L+H^L_{env}+H_I^L$ 
where $S=-\frac{1}{\sqrt{N}}\sum_{i,k}g_{k} (n_i-\frac{1}{2}) (a_{i,k} - a_{i,k}^{\dagger})$. 
Then, the components of $H^L$ (i.e., the systems part $H_s^{L}$, the environment part $H^L_{env}$, and the interaction part $H^L_I$) are expressed as
\begin{eqnarray}
\!\!\!\! H_s^{L}&=&\varepsilon_{1} (n_{1}-\frac{1}{2})+\varepsilon_{2} (n_{2}-\frac{1}{2})
+ J_\lVert (n_{1}-\frac{1}{2})(n_{2}-\frac{1}{2})
\nonumber \\ && 
- \frac{J_\perp e^{-\frac{1}{N}\sum_{k}g_{k}^2}}{2} (b_{1}^{\dagger}b_{2}+b_{2}^{\dagger}b_{1}) ,
\end{eqnarray}
\begin{eqnarray}
H^L_{env}=\sum_{i,k} \omega_{k} a_{i,k}^{\dagger} a_{i,k} ,
\end{eqnarray}
and
\begin{eqnarray}
 H_I^L = -\frac{1}{2} [J_\perp ^+ b_{1}^{\dagger}b_{2} + J_\perp ^- b_{2}^{\dagger}b_{1} ], 
\end{eqnarray}
where
\begin{eqnarray}
\!\! J_\perp ^{\pm} &=& J_\perp e^{\pm \frac{1}{\sqrt{N}}\sum_{k}~ g_k [(a_{2,k} - a_{2,k}^{\dagger})-(a_{1,k} - a_{1,k}^{\dagger})]} - 
 J_\perp e^{-\frac{1}{{N}}\sum_{k}g_k^2} .
 \nonumber \\
\end{eqnarray}
Now, we use the non-Markovian master Eq. (\ref{mas}) to study the dynamics of the reduced density matrix. We calculate below the matrix element 
$_{ph}\langle \{0_1^k\},\{0_2^k\}|H_I^{L}|\{m_1^k\},\{m_2^k\}\rangle_{ph}$ with $m_1^k$ and $m_2^k$ being the 
occupation numbers of the k-th mode phonons at 
site 1 and site 2, respectively.
\begin{eqnarray}
 \!\! &&_{ph}\langle \{0_1^k\},\{0_2^k\}|H_I^{L}|\{m_1^k\},\{m_2^k\}\rangle_{ph} \nonumber \\
&=&-\frac{J_\perp}{2} e^{-\frac{1}{{N}}\sum_{k}g_k^2} \Bigg(\prod_k \frac{(\frac{g_k}{\sqrt{N}})^{(m_1^k+m_2^k)}}{\sqrt{m_1^k!m_2^k!}}
 \Bigg)
 \nonumber \\
 && \times(-1)^{\sum_{k} m_1^k} \big[b_{1}^{\dagger}b_{2}+(-1)^{\sum_{k} (m_1^k-m_2^k)}b_{2}^{\dagger}b_{1}\big].
\nonumber \\
\end{eqnarray}
Using the above result and Eqs. (\ref{basis1}) and (\ref{basis2})
 \Big [with $\kappa$ replaced by $\bar{\kappa} \equiv-\frac{J_{\perp} e^{-\frac{1}{N}\sum_{k}g_{k}^2}}{2}$ \Big], we calculate the four terms in 
the master Eq. (\ref{mas}); in the regime where 
$\Delta\varepsilon\gg J_{\perp} e^{-\frac{1}{N}\sum_{k}g_{k}^2}$, we can write the differential equation for $\langle 10 | \tilde{\rho}_s(t)|01\rangle$ 
to be 
\begin{widetext}
\begin{eqnarray}
\frac{d \langle 10 | \tilde{\rho}_s(t)|01\rangle }{dt}&=&-\bar{\kappa}^2 \sum^{\prime}_{\{n^{k}_1\},\{n^k_2\}}\bar{C}_{n} 
\int^{t}_{0} d\tau\Bigg[\langle 10 | \tilde{\rho}_s(t)|01\rangle
 \bigg( e^{i(\bar{\omega}_n+\Delta \varepsilon)T}+e^{-i(\bar{\omega}_n-\Delta \varepsilon)T}\bigg) \nonumber \\
&&~~~~~~~~~~~~~~~~~~~~~~- \langle 01 | \tilde{\rho}_s(t)|10\rangle (-1)^{\sum_{k}(n_{1}^{k}-n_{2}^{k})} e^{2i\Delta \varepsilon t}\bigg(  e^{i(\bar{\omega}_n-\Delta 
\varepsilon)T}+e^{-i(\bar{\omega}_n+\Delta \varepsilon)T} \bigg)\Bigg] .
\label{offmulti}
\end{eqnarray}
The corresponding complex conjugate equation would describe the dynamics 
for $\langle 01 | \tilde{\rho}_s(t)|10\rangle$. Here,  we have defined $\bar{\omega}_{{n}}\equiv\sum_{k} \omega_k (n_1^k+n_2^k)$,
 $\bar{C}_n\equiv\prod_k \frac{(\frac{g_k}{\sqrt{N}})^{2(n_1^k+n_2^k)}}{{n_1^k!n_2^k!}}$, 
$T \equiv t-\tau$, and $\sum^{\prime}_{\{n^{k}_1\},\{n^k_2\}}$ as the sum over 
all combinations of $\{n^{k}_1 \}$  and $\{n^k_2\}$ excluding the case when $\{n_1^k\}=\{0_1^k\}$ and $\{n_2^k\}=\{0_2^k\}$. 
Similarly, one can obtain the following differential equation for $\langle 10 | \tilde{\rho}_s(t)|10\rangle$: 
\begin{eqnarray}
 \frac{d \langle 10 | \tilde{\rho}_s(t)|10\rangle }{dt}&=&-\bar{\kappa}^2\sum^{\prime}_{\{n^{k}_1\},\{n^k_2\}}\bar{C}_{n} \int^{t}_{0} d\tau\Bigg[2\langle 10 | \tilde{\rho}_s(t)|10\rangle
 {\rm cos}(\Delta\varepsilon T) \bigg( e^{i\bar{\omega}_nT}+e^{-i\bar{\omega}_nT}\bigg) \nonumber \\
 &&~~~~~~~~~~~~~~~~~~~~~~~-\bigg(e^{i(\bar{\omega}_n+\Delta \varepsilon)T}+e^{-i(\bar{\omega}_n+\Delta \varepsilon)T} \bigg) \Bigg].
 \label{diamulti}
\end{eqnarray}
Now, to get a closed form of the coefficients in Eqs. (\ref{offmulti}) and (\ref{diamulti}), we write
\begin{eqnarray}
 \sum^{\prime}_{\{n^{k}_1\},\{n^k_2\}} \bar{C}_n e^{\pm i\bar{\omega}_n T}
 &=&\prod_{k} \Bigg( \sum_{n^{k}_1} \frac{\Big(\frac{g_{k}^2}{N}\Big)^{n^{k}_1}} {n^{k}_{1}!} e^{\pm i\omega_k n^{k}_1 T}
 \sum_{n^{k}_2} \frac{\Big(\frac{g_{k}^2}{N}\Big)^{n^{k}_2}} {n^{k}_{2}!} e^{\pm i\omega_k n^{k}_2 T}\Bigg)-1 \nonumber \\
 &&={\rm exp} \bigg[ \sum_{k} 2 \frac{g^{2}_{k}}{N} e^{\pm i\omega_k T}\bigg]-1 \nonumber \\
 &&={\rm exp} \bigg[\frac{2}{N\pi}\int_0^{\infty}d \omega \frac{J(\omega)}{\omega^2}e^{\pm i\omega T} \bigg] -1,
 \label{density0}
\end{eqnarray}
where the spectral function of the phonon bath $J(\omega)=\pi\sum_{k}g_k^2 \omega_k^2 \delta (\omega-\omega_k)$ characterizes the HCB-phonon coupling for 
different phonon-frequency modes. Using the above expression, we can write the differential Eqs. (\ref{offmulti}) and (\ref{diamulti}) as 
\begin{eqnarray}
&& \!\!\!\!\!\!\!\!\!\!\!\! \frac{d \langle 10 | \tilde{\rho}_s(t)|01\rangle }{dt} \nonumber \\
&&\!\!\!\!\!\! =-2\bar{\kappa}^2 \int_0^t d \tau \Bigg[\langle 10 | \tilde{\rho}_s(t)|01\rangle
 e^{i\Delta \varepsilon T} \bigg( {\rm exp} \Big[\frac{2}{N\pi}\int_0^{\infty}d \omega \frac{J(\omega)}{\omega^2}{\rm cos}(\omega T) \Big]
 {\rm cos}\Big[\frac{2}{N\pi}\int_0^{\infty}d \omega \frac{J(\omega)}{\omega^2}{\rm sin}(\omega T) \Big]-1  \bigg) \nonumber \\
&&~~~~~~~~~~~~~~~~ -\langle 01 | \tilde{\rho}_s(t)|10\rangle e^{i\Delta \varepsilon(t+\tau)} \bigg( {\rm exp} \Big[-\frac{2}{N\pi}\int_0^{\infty}d \omega \frac{J(\omega)}{\omega^2}{\rm cos}(\omega T) \Big]
 {\rm cos}\Big[\frac{2}{N\pi}\int_0^{\infty}d \omega \frac{J(\omega)}{\omega^2}{\rm sin}(\omega T) \Big]-1  \bigg)\Bigg] ,
\label{denoff}
\end{eqnarray}
and 
\begin{eqnarray}
&& \!\!\!\!\!\!\!\!\!\!\!\! \frac{d \langle 10 | \tilde{\rho}_s(t)|10\rangle }{dt} \nonumber \\
&&\!\!\!\!\!\! =-2\bar{\kappa}^2\int_0^t d \tau \Bigg[ 2\langle 10 | \tilde{\rho}_s(t)|10\rangle
 {\rm cos}(\Delta \varepsilon T)\bigg( {\rm exp} \Big[\frac{2}{N\pi}\int_0^{\infty}d \omega \frac{J(\omega)}{\omega^2}{\rm cos}(\omega T) \Big]
 {\rm cos}\Big[\frac{2}{N\pi}\int_0^{\infty}d \omega \frac{J(\omega)}{\omega^2}{\rm sin}(\omega T) \Big]-1 \bigg)\nonumber \\
&&~~~~~~~~~~~~~~~~ -\bigg( {\rm exp} \Big[\frac{2}{N\pi}\int_0^{\infty}d \omega \frac{J(\omega)}{\omega^2}{\rm cos}(\omega T) \Big] 
 {\rm cos}\Big[\Delta \varepsilon T+\frac{2}{N\pi}\int_0^{\infty}d \omega \frac{J(\omega)}{\omega^2}{\rm sin}(\omega T) \Big]- 
{\rm cos}(\Delta \varepsilon T)\bigg)
 \Bigg].
\label{dendia}
\end{eqnarray}
The differential Eq. (\ref{dendia}) can be solved analytically and the solution is written as 
\begin{eqnarray}
 \langle 10 | {\rho}_s(t)|10\rangle&=& e^{-\int_{0}^{t} dt^{\prime} A(t^{\prime})} \Bigg( \langle 10 | {\rho}_s(0)|10\rangle 
 +\int_{0}^{t}dt^{\prime} B(t^{\prime}) e^{\int_{0}^{t^{\prime}} dt^{\prime\prime} A(t^{\prime\prime})}\Bigg),
 \label{multisol}
 \end{eqnarray}
where 
\begin{eqnarray}
A(t)=4\bar{\kappa}^2 \int_0^t d \tau ~{\rm cos}(\Delta \varepsilon T)\Bigg( {\rm exp} 
\bigg[ \frac{2}{N\pi}\int_0^{\infty}d \omega \frac{J(\omega)}{\omega^2}{\rm cos}(\omega T) \bigg]
 {\rm cos}\bigg[ \frac{2}{N\pi}\int_0^{\infty}d \omega \frac{J(\omega)}{\omega^2}{\rm sin}(\omega T) \bigg]-1 \Bigg) ,
\end{eqnarray}
 and 
\begin{eqnarray}
 B(t)=2\bar{\kappa}^2 \int_0^t d \tau \Bigg( {\rm exp} \bigg[\frac{2}{N\pi}\int_0^{\infty}d \omega \frac{J(\omega)}{\omega^2}{\rm cos}(\omega T) \bigg] 
 {\rm cos}\bigg[\Delta \varepsilon T+\frac{2}{N\pi}\int_0^{\infty}d \omega \frac{J(\omega)}{\omega^2}{\rm sin}(\omega T) \bigg]- 
{\rm cos}(\Delta \varepsilon T)\Bigg) .
\end{eqnarray}
\end{widetext}
In principle,  $J(\omega)$ can assume a variety of forms based on the nature of the phonon bath; however, for simplicity, we use a continuous 
uniform distribution of phonon frequencies within a small frequency window characterized by an upper cutoff $\omega_u$ and a lower cutoff $\omega_l$. 
The density of states for Einstein phonons is described by ${\rm D}(\omega_k)=N \delta (\omega_k-\omega_0)$ where  $\omega_0$ 
is a fixed frequency and $N$ is the  number of phonon modes. Moreover, we consider a weak k-dependence of the coupling strength $g_k$ and write 
\begin{eqnarray}
  {\rm D}(\omega_k)g_k^2=N \delta (\omega_k-\omega_0)g^2.
\label{einst}
\end{eqnarray}
Here we should mention that in systems such as the manganites (where the carriers are coupled predominantly only to optical phonons),
 the weak k-dependence of $g_k$ is quite valid. 
Following Eq. (\ref{einst}) we make a simple generalization of Einstein model and replace the Dirac delta function by a box function of width 
$(\omega_u -\omega_l)$ and height $\frac{1}{(\omega_u -\omega_l)}$
\begin{eqnarray}
  {\rm D}(\omega_k)g_k^2&=& g^2 \frac{N}{\omega_u - \omega_l} \Theta (\omega_k - \omega_l) \Theta(\omega_u - \omega_k) ,
 \label{density}
\end{eqnarray}
where $\Theta(\omega)$ is the unit step function.
With the above form for the density of states, we calculate the following:
 \begin{eqnarray}
\frac{1}{N\pi}\int_0^\infty \frac{J(\omega)}{\omega^2} d\omega &=& \frac{1}{N}\sum_k g_k^2 \nonumber \\
&=&\frac{1}{N} \int_0^\infty d\omega_k {\rm D}(\omega_k) g_k^2 \nonumber \\
&=&\int_{\omega_l}^{\omega_u} d\omega_k \frac{g^2}{\omega_u - \omega_l}\nonumber \\
&=&g^2 ,
\label{prefactor}
 \end{eqnarray}
 \begin{eqnarray}
\!\!\!\!\!\! \frac{1}{N\pi}\int_0^\infty \frac{J(\omega)}{\omega^2} \cos\omega T d\omega &=& \frac{1}{N}\sum_k g_k^2 \cos\omega_k T\nonumber \\
 &=&\frac{1}{N} \int_0^\infty d\omega_k {\rm D}(\omega_k) g_k^2 \cos\omega_k T\nonumber \\
 &=&\frac{g^2}{(\omega_u - \omega_l)T} (\sin\omega_u T - \sin\omega_l T)\nonumber \\
 &=&\frac{2g^2}{(\omega_u - \omega_l)T} \cos\Big[\frac{(\omega_u+\omega_l)T}{2}\Big]\nonumber \\
 &&~~~~~~~~~~~~~~\times \sin\Big [\frac{(\omega_u-\omega_l)T}{2}\Big] ,
\nonumber \\
 \label{cos}
\end{eqnarray}
and
\begin{eqnarray}
\!\!\!\!  \frac{1}{N\pi}\int_0^\infty \frac{J(\omega)}{\omega^2} \sin\omega T d\omega &=& \frac{1}{N}\sum_k g_k^2 \sin\omega_k T\nonumber \\
 &=&\frac{1}{N} \int_0^\infty d\omega_k {\rm D}(\omega_k) g_k^2 \sin\omega_k T\nonumber \\
 &=&\frac{g^2}{(\omega_u - \omega_l)T} (\cos\omega_l T - \cos\omega_u T) \nonumber \\
 &=&\frac{2g^2}{(\omega_u - \omega_l)T} \sin \Big [\frac{(\omega_u+\omega_l)T}{2} \Big] \nonumber \\
 &&~~~~~~~~~~~~~~\times \sin \Big [\frac{(\omega_u-\omega_l)T}{2}\Big] .
\nonumber \\
 \label{sin}
\end{eqnarray}
Using the above integrals, we solve the differential Eq. (\ref{denoff}) numerically and plot the coherence factor ${\rm C(t)}$ in Fig. (\ref{fig11}). 
Here, unlike the single-mode case, we have a continuum of phonon frequencies due to which the various harmonics in Eq. (\ref{offmulti})
do not all rephase at the same time leading to destructive interference, i.e., an irreversible decay of
${\rm C(t)}$.

 To explain the periodicity of the plot in Fig. \ref{fig11}, 
we rewrite Eq. (\ref{offmulti}) as 
\begin{widetext}
\begin{eqnarray}
\!\!\!\!\!\!\!\!\!\!\!\!  \frac{d \langle 10 | \tilde{\rho}_s(t)|01\rangle }{dt}&=&-i\bar{\kappa}^2 \!\!\sum^{\prime}_{\{n^k_1\},\{n^k_2\}}
\!\! \bar{C}_{n} \Bigg[\langle 10 | \tilde{\rho}_s(t)|01\rangle
  \bigg(\frac{e^{-i(\bar{\omega}_n -\Delta\varepsilon)t}}{\bar{\omega}_n -\Delta\varepsilon} -\frac{e^{i(\bar{\omega}_n +\Delta\varepsilon)t}}{\bar{\omega}_n+\Delta\varepsilon}
  -\frac{2\Delta\varepsilon}{\bar{\omega}^{2}_n-\Delta\varepsilon^2}\bigg) \nonumber \\ 
 &&~~~~~~~~~~~~~~~~~~~~ + \langle 01 | \tilde{\rho}_s(t)|10\rangle \big(-1\big)^{\sum_{k}(n_{1}^{k}-n_{2}^{k})}
 e^{2i\Delta \varepsilon t} \bigg( \frac{e^{i(\bar{\omega}_n -\Delta\varepsilon)t}}{\bar{\omega}_n -
  \Delta\varepsilon} -\frac{e^{-i(\bar{\omega}_n +\Delta\varepsilon)t}}{\bar{\omega}_n+\Delta\varepsilon}
  -\frac{2\Delta\varepsilon}{\bar{\omega}^{2}_n-\Delta\varepsilon^2}    \bigg)  \Bigg].
  \label{offk}
\end{eqnarray}
\end{widetext}

\begin{figure}[b]
\centerline{\includegraphics[angle=90,angle=90,angle=90,width=3.2in,height=3.5in]{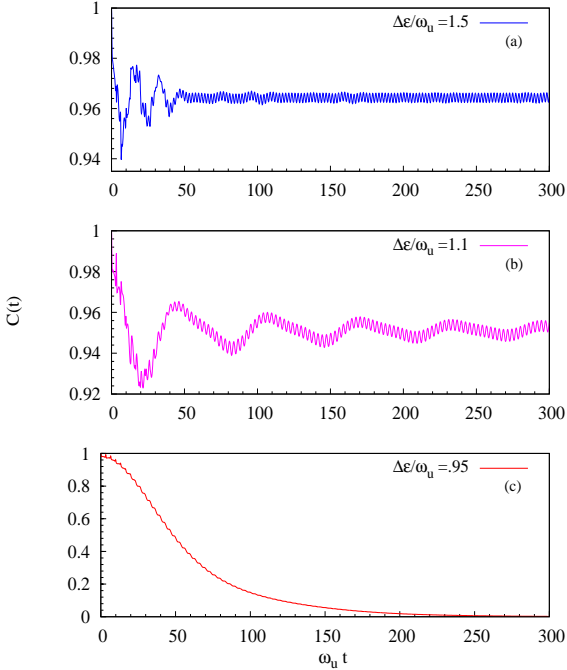}}
\caption{Time dependence of ${\rm C(t)}$ for 
$\frac{J_{\perp}}{\omega_{u}}=1.0$, $\frac{\omega_l}{\omega_u}=0.9$, $g=2.0$ and different values of $\frac{\Delta \varepsilon}{\omega_{u}}$
leading to different scenarios.}
\label{fig11}
\end{figure}

\noindent It is of interest to note that the structures of Eqs. (\ref{offeqn}) and (\ref{offk}) are very similar; hence the
explanations that were offered in the single-mode case, 
for the period and amplitude of oscillations as well as for the equilibrium values of ${\rm C(t)}$, hold also for the multimode case.
For the circumstance in Fig. \ref{fig11} (b), the contribution from the phonon state $\omega_u$ dominates because it is
the frequency that is closest to $\Delta \varepsilon$ and $\Delta \varepsilon -\omega_u$ ($<< \omega_u$) is comparable to the 
width of the allowed-frequency window $\omega_u-\omega_l$.   
Then, the period of  oscillation in Eq. (\ref{offk}) can be obtained approximately from the case
$|(\bar{\omega}_n -\Delta\varepsilon)t|= |(\omega_u -\Delta\varepsilon)t|=0.1\omega_u t$; 
thus the period is approximately $20\pi/\omega_u$.  Furthermore, since 
$\Delta \varepsilon$ is close to $\bar{\omega}_n$,
only a few frequencies contribute to the destructive interference leading to  a gradual decay of the amplitude
 of oscillation in Fig. \ref{fig11} (b). 

For the situation where $\Delta\varepsilon$ equals  any of the phonon eigenenergies $\bar{\omega}_n$
[such as in Fig. \ref{fig11} (c)], there is a complete decay  
of coherence due to strong exchange of energy.
When $\Delta\varepsilon$ is away from $\bar{\omega}_n$ [which is the case in Fig. \ref{fig11} (a)], there are a number of dominant phonon states 
having comparable contributions and these states interfere destructively resulting in a quick decay of amplitude.

\begin{figure}[t]
\centerline{\includegraphics[angle=90,angle=90,angle=90,width=3.2in,height=3.5in]{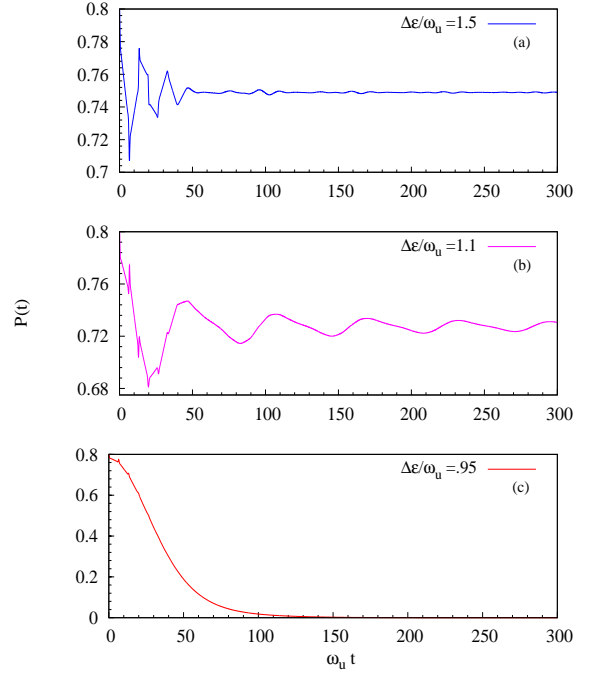}}
\caption{Time dependence of ${\rm P(t)}$ for 
$\frac{J_{\perp}}{\omega_{u}}=1.0$, ${\rm P(0) = 0.8}$, $\frac{\omega_l}{\omega_u}=0.9$, $g=2.0$ and different values of $\frac{\Delta \varepsilon}{\omega_{u}}$.}
\label{fig12}
\end{figure}

Next, we study the population ${\rm P(t)}$ of the excited state $|10\rangle$ and depict its variation in Fig. \ref{fig12}. When the excited 
state is initially largely populated (such as in Fig. \ref{fig12} where $\langle 10 | {\rho}_s(0)|10\rangle=0.8$), 
 the behavior of
${\rm  P(t)}$ is mainly dictated by the 
homogeneous part of the solution given in Eq. (\ref{diamulti}). To explain the behavior of Fig. \ref{fig12}, we rewrite Eq. (\ref{diamulti}) as 
follows:
\begin{widetext}
\begin{eqnarray}
\!\!\!\!\!\!  \frac{d \langle 10 | \tilde{\rho}_s(t)|10\rangle }{dt}&=&-2\bar{\kappa}^2 \!\! \sum^{\prime}_{\{n^k_1\},\{n^k_2\}} \!\! \bar{C}_{n} 
\Bigg[\langle 10 | \tilde{\rho}_s(t)|10\rangle
  \bigg(\frac{\sin (\bar{\omega}_n +\Delta\varepsilon)t}{\bar{\omega}_n +\Delta\varepsilon} 
+\frac{\sin (\bar{\omega}_n -\Delta\varepsilon)t}{\bar{\omega}_n-\Delta\varepsilon} \bigg)
-\frac{\sin (\bar{\omega}_n +\Delta\varepsilon)t}{\bar{\omega}_n +\Delta\varepsilon} \Bigg ]   .
  \label{diak}
\end{eqnarray}
\end{widetext}

Since the structures of Eqs. (\ref{diageqn}) and (\ref{diak}) are very similar, we expect that the single-mode and multimode cases
will have similar justifications
for the period and amplitude of oscillations as well as for the equilibrium values of ${\rm P(t)}$.
The cases of $\Delta \varepsilon$ considered in Fig. \ref{fig12} are the same
as those studied in Fig. \ref{fig11}; furthermore, the same explanations hold for the period and decay of oscillations in these two figures.

\begin{figure}
\centerline{\includegraphics[angle=90,angle=90,angle=90,width=3.2in,height=2.5in]{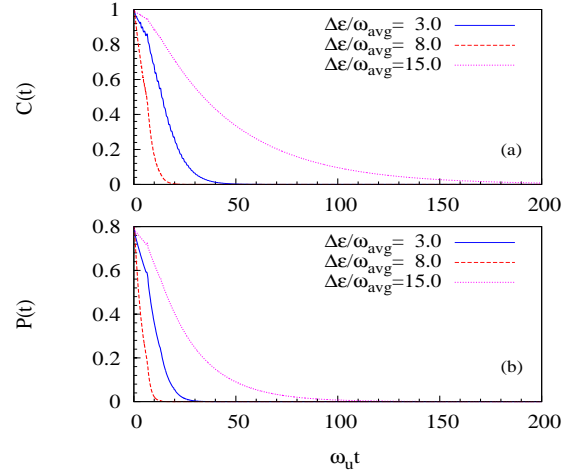}}
\caption{
Depiction of complete decay of (a) ${\rm C(t)}$ and (b) ${\rm P(t)}$ [with ${\rm P(0) = 0.8}$] 
 at different integer values of $\frac{\Delta \varepsilon}{\omega_{\rm avg}}$
when $\frac{J_{\perp}}{\omega_{u}}=1.0$ and  $g=2.0$.
The maximum decay occurs when $\Delta \varepsilon = 2 g^2 \omega_{\rm avg}$.}
\label{fig13}
\end{figure}

Lastly, we elucidate through Fig. \ref{fig13} the multimode cases 
\Big [of $\Delta \varepsilon$ being an integer multiple of $\omega_{\rm avg}\equiv \frac{1}{N}\sum_k \omega_k =(\omega_u+\omega_l)/2$ \Big ]
when the coherence ${\rm C(t)}$ and the excited state population ${\rm P(t)}$
undergo complete decay. Similar to the single-mode case [depicted in Figs. \ref{fig3}(c) and \ref{fig4}(c)],
 here too the maximum decay of both ${\rm C(t)}$ and ${\rm P(t)}$ occurs when
$\Delta \varepsilon$ is equal to twice the polaron energy $\frac{1}{N}\sum_k g^2_k \omega_k$ (i.e., $ 2 g^2 \omega_{\rm avg}$ for our density of states).

\section{Discussion and Conclusions}
In this work we  considered two models of HCB, characterized by strong HCB-phonon coupling and antiadiabaticity, 
with the initial state involving no correlation between the system and the environment in the polaronic frame of reference (where the interaction term is weak).
These two models are a generalization of the systems studied in Ref. \onlinecite {polarondyn}.  
(i) We have shown that an infinite-range HCB model subject to Markovian dynamics does not show decoherence or decay of excited state population.
It is the long-range nature of the Hamiltonian and  the negligible renormalized hopping  (compared to the phonon energy)
 that preserve the eigenstates of the system coupled to strong local optical
phonons and ensure decoherence free dynamics. (ii) The more realistic case, of non-Markovian dynamics and site energy differences
being non-negligible compared to phonon energy,  has been analyzed for an amenable two-site system.
When the site-energy difference is not close to a phonon eigenenergy, the amount of decoherence and decay of excited state
are both quite small and close to the Markovian results; whereas, decoherence and decay become prominent as system energy approaches a phonon 
eigenenergy.

{We should mention that the 
approximate results obtained in the previous section (by neglecting $\frac{|\kappa|}{|\Delta\varepsilon|}$ compared to 1)
are very close to the results obtained without any approximation (i.e., using the full expressions obtained in the appendix).
 Moreover, for the numerical results in the previous section, we used fourth-order Runge-Kutta for solving differential equations 
and Gaussian quadrature  for numerical integrations.}

We will now make a few general remarks regarding the range of hopping in a multi-site case.
In contrast to our long-range  model involving distance-independent hopping of HCBs,
if we were to consider a chain with 
 nearest-neighbor (NN) hopping 
\Big [of the type $\sum_{i}\{\frac{-J_{\perp}}{2}( b^{\dagger}_i  b_{i+1} +{\rm H.c.}) + J_{\lVert} (n_i - \frac{1}{2})(n_{i+1} -\frac{1}{2})\}$  
(with $J_{\parallel} = 0$)\Big ]  and strongly couple the HCBs
to local phonons \Big[by introducing the additional terms
 $g \omega \sum_i (n_i- \frac{1}{2}) (a^{\dagger}_i + a_i) 
+  \omega \sum_i  a^{\dagger}_i a_i$\Big], we get decoherence for the case of half-filling. The NN-hopping system transits from a superfluid,
with large values of the off-diagonal density matrix terms 
\Big [i.e.,  $<b^{\dagger}_i \sum_{j \neq i} b_j>$ = Bose-Einstein condensate occupation number $n_0$
 $\propto \sqrt{N}$ (see Ref. \onlinecite{srsypbl2})\Big],
to a  charge-density-wave state with
significantly diminished off-diagonal density matrix terms 
\Big( $<b^{\dagger}_i \sum_{j \neq i} b_j>$\Big). The above analysis can be
mapped (through a HCB-to-spin transformation and then a Wigner-Jordan transformation) on to  the analysis in Ref. \onlinecite{sdadys}
dealing with the transition from a  Luttinger liquid to
a charge-density-wave.
Furthermore, the eigenstates of the effective Hamiltonian are not the same as
the original 
 NN-hopping model
(for the $J_{\parallel} =0$ case);
the effective Hamiltonian contains additional next-nearest-neighbor hopping
terms $b^{\dagger}_{i+1}b_{i-1}$ and additional NN repulsion
terms $n_{i+1}n_i$
 that are not present in the original 
Hamiltonian [see Eqs. (4) and (5) in Ref. \onlinecite{sdadys}].
It is important to note that the infinite-range
HCB model gives decoherence free behavior whereas
the NN HCB model does not; thus, the range of interaction determines
the decoherence of the system even when $J^{\star}e^{-g^2} \ll \omega$.
}

Although the 
analysis in this paper is valid for optical phonons, 
it can also accommodate acoustic phonons in small systems because 
{the smallest 
 wavevector, 
for a system with fixed boundaries, is}
inversely proportional to the system size; hence, for a small system $\Delta \varepsilon$ can be different from the eigenenergies of acoustic phonons.
Furthermore, based on our study of a two-site system,
our inferences can be extrapolated to a many-site situation, namely: as long as the various site-energy differences in a multi-site
(as well as a many-body) case are away from the phonon eigenenergies, the decoherence will be small.
This also implies that, to realize large coherence, the phonon density-of-states should vanish below a lower cut-off frequency. 
 Lastly, the above analysis is valid in the regime $k_B T /\omega \ll 1$; the finite temperature
case  $k_B T /\omega \gtrsim 1$ needs additional extensive considerations
and will be dealt with elsewhere.

The 
two-site case can be thought of as a system of an acceptor and a donor with different site energies (due to defects, impurities, etc.);
the dynamics of population transfer between them as well as 
the two-site coherence are important for understanding many physical systems such as a double quantum dot (DQD) acting as a qubit
 for quantum computation \cite{polarondyn}, 
artificial light-harvesters, etc. 
An oxide- (i.e., manganite-) based DQD \cite{polarondyn} can serve as a charge qubit and meet the demands of miniaturization
as it 
has very small decoherence (compared to a semiconductor DQD) and its size can also be much smaller
than a semiconductor DQD \cite{semidqd}.
Additionally, understanding the high ($>90\%$) quantum efficiency of energy transport between various chlorophyll molecules in photosynthesis
 is important to design artificial solar energy applications;
minimizing the decoherence in an interacting many-spin system coupled to the environment is quite useful 
for developing quantum computer architecture.
 Our analysis of a many-body HCB model with  Markovian dynamics is a step to meet the above ends.   

\section{Acknowledgments}
{ One of the authors (S. Y.) would like to thank 
G. Baskaran, P. B. Littlewood, R. Simon, S. Ghosh, and S. Reja for valuable discussions.}

\appendix
\section{}
\begin{widetext}
Here, we calculate the four terms (on the right hand side) of the integrand in the following master equation obtained from Eq. (\ref{mas}) in the main text.
\begin{eqnarray}
\frac{d \langle 10| \tilde{\rho}_s(t)|01\rangle}{dt} &=& - \sum_{n} \int_0^t d\tau \bigg[~ \langle 10|_{ph}\langle 0 |
 \tilde{H}_I^L(t)| n \rangle_{ph}~ _{ph}\langle n| \tilde{H}_I^L(\tau)| 0 \rangle_{ph} \tilde{\rho}_s(t) |01\rangle  
\nonumber \\ 
 &&~~~~~~~~~~~~~~~~~~~
 - \langle 10|{~_{ph}}\langle n| \tilde{H}_I^L(t)|0\rangle_{ph}\tilde{\rho}_s(t) {_{ph}}\langle 0 | \tilde{H}_I^L(\tau) |n\rangle_{ph}|01\rangle \nonumber \\
 && ~~~~~~~~~~~~~~~~~~~- \langle 10|{~_{ph}}\langle n| \tilde{H}_I^L(\tau)| 0\rangle_{ph} \tilde{\rho}_s(t) {_{ph}}\langle 0| \tilde{H}_I^L(t) |n\rangle_{ph}|01\rangle 
 \nonumber \\
&& ~~~~~~~~~~~~~~~~~~~
 + 
 \langle 10|\tilde{\rho}_s(t)  _{ph}\langle 0|
 \tilde{H}_I^L(\tau)| n \rangle_{ph}~ _{ph}\langle n | \tilde{H}_I^L(t)| 0\rangle_{ph}|01\rangle  
 \bigg ] .
\label{masele}
\end{eqnarray}
The First term is evaluated as follows:
\begin{eqnarray}
 &&\sum^{\prime}_{n} \langle 10|_{ph}\langle 0 |
 \tilde{H}_I^L(t)| n \rangle_{ph}~ _{ph}\langle n| \tilde{H}_I^L(\tau)| 0 \rangle_{ph} \tilde{\rho}_s(t) |01\rangle \nonumber \\
 &&= \sum^{\prime}_{n}  \langle 10| e^{iH^{L}_{s} t} {_{ph}}\langle 0 |
 {H}_I^L| n \rangle_{ph} e^{-iH^{L}_{s} (t-\tau)} {_{ph}}\langle n| {H}_I^L| 0 \rangle_{ph} e^{-iH^{L}_{s} \tau}\tilde{\rho}_s(t) |01\rangle
 e^{-i\omega_n (t-\tau)} \nonumber \\
 &&=\kappa^2 \sum^{\prime}_{n} C_n \langle 10|  e^{iH^{L}_{s} t} (b_{1}^{\dagger}b_{2}+(-1)^{(n_1+n_2)} b_{2}^{\dagger}b_{1})e^{-iH^{L}_{s} (t-\tau)}
 \Big ( b_{2}^{\dagger}b_{1}+(-1)^{n_1+n_2} b_{1}^{\dagger}b_{2} \Big ) e^{-iH^{L}_{s} \tau}\tilde{\rho}_s(t) |01\rangle
 e^{-i\omega_n (t-\tau)} \nonumber \\
 &&=\kappa^2 \sum^{\prime}_{n} C_n \bigg[ \langle 10|\tilde{\rho}_s(t) |01\rangle \Big\{p(t)p(t-\tau)p^*(\tau)+(-1)^{(n_1+n_2)} \kappa^2 q(t)q(t-\tau)
p^*(\tau)\nonumber \\
&&~~~~~~~~~~~~~~~~~~~~~~~~~~~~~~~~~~~+ \kappa^2 q(t) p^*(t-\tau)q(\tau)-(-1)^{(n_1+n_2)}\kappa^2 p(t)q(t-\tau)q(\tau)\Big\} \nonumber \\
&&~~~~~~~~~~~~~~~~-i\kappa \langle 01|\tilde{\rho}_s(t) |01\rangle \Big\{p(t)p(t-\tau)q(\tau)+\kappa^2 (-1)^{(n_1+n_2)}q(t)q(t-\tau)q(\tau) \nonumber \\
&&~~~~~~~~~~~~~~~~~~~~~~~~~~~~~~~~~~~~~ -q(t)p^* (t-\tau) p(\tau)+(-1)^{(n_1+n_2)} p(t)q(t-\tau)p(\tau)\Big\}\bigg] e^{-i\omega_n (t-\tau)} ~.
\label{1st}
\end{eqnarray}

The second term is given by\\
\begin{eqnarray}
 &&\sum^{\prime}_{n}\langle 10|{~_{ph}}\langle n| \tilde{H}_I^L(t)|0\rangle_{ph}\tilde{\rho}_s(t) 
 {_{ph}}\langle 0 | \tilde{H}_I^L(\tau) |n\rangle_{ph}|01\rangle \nonumber \\
 &&=\sum^{\prime}_{n} \langle 10| e^{iH^{L}_{s} t}{_{ph}}\langle n |
 {H}_I^L| 0 \rangle_{ph} e^{-iH^{L}_{s} t}\tilde{\rho}_s(t)e^{iH^{L}_{s} \tau}\langle 0 |
 {H}_I^L| n \rangle_{ph} e^{-iH^{L}_{s} \tau}|01\rangle e^{i\omega_n (t-\tau)}\nonumber \\
 &&=\kappa^2 \sum^{\prime}_{n} C_{n} \langle 10|e^{iH^{L}_{s} t} \Big(b_{2}^{\dagger}b_{1}+(-1)^{n_1+n_2} b_{1}^{\dagger}b_{2}\Big)
 e^{-iH^{L}_{s} t}\tilde{\rho}_s(t)e^{iH^{L}_{s} \tau}\Big (b_{1}^{\dagger}b_{2}+(-1)^{(n_1+n_2)} b_{2}^{\dagger}b_{1}\Big)e^{-iH^{L}_{s} \tau}
 |01\rangle e^{i\omega_n (t-\tau)}\nonumber \\
 &&=\kappa^2 \sum^{\prime}_{n} C_{n} \bigg[\kappa^2\langle 10|\tilde{\rho}_s(t) |01\rangle  \Big\{(-1)^{(n_1+n_2)}q(t)q(\tau)p(t)p(\tau)
 -q(t)p^* (\tau) p(t)q(\tau) \nonumber \\
 &&~~~~~~~~~~~~~~~~~~~~~~~~~~~~~~~~~~~~~~-p^*(t)q(\tau)q(t)p(\tau)+ (-1)^{(n_1+n_2)} p^* (t)p^* (\tau)q(t)q(\tau)\Big\}\nonumber \\
&& ~~~~~~~~~~~~~~~~+\langle 01|\tilde{\rho}_s(t) |10\rangle \Big\{(-1)^{(n_1+n_2)}p^2(t)p^2(\tau)+\kappa^2 p^2(t) q^2 (\tau)\nonumber \\
 &&~~~~~~~~~~~~~~~~~~~~~~~~~~~~~~~~~~~~~~+\kappa^2 q^2(t) p^2 (\tau)+(-1)^{(n_1+n_2)}\kappa^4 q^2(t) q^2 (\tau)\Big\}\nonumber \\
 &&~~~~~~~~~~~~~~~~+i\kappa\langle 10|\tilde{\rho}_s(t) |10\rangle  \Big\{-(-1)^{(n_1+n_2)}q(t)p^2(\tau)p(t)-\kappa^2 q(t)q^2(\tau)p(t)\nonumber\\
 &&~~~~~~~~~~~~~~~~~~~~~~~~~~~~~~~~~~~~~+p^*(t)p^2(\tau)q(t)+(-1)^{(n_1+n_2)}\kappa^2 p^*(t)q^2(\tau)q(t)\Big\}\nonumber \\
 &&~~~~~~~~~~~~~~~~+i\kappa\langle 01|\tilde{\rho}_s(t) |01\rangle \Big\{(-1)^{(n_1+n_2)}p^2(t)q(\tau)p(\tau)-p^*(\tau)p^2(t) q(\tau) \nonumber \\
 &&~~~~~~~~~~~~~~~~~~~~~~~~~~~~~~~~~~~~~+\kappa^2 q^2(t)q(\tau)p(\tau)-(-1)^{(n_1+n_2)}\kappa^2q^2 (t) p^*(\tau)q(\tau)\Big\}\bigg]e^{i\omega_n (t-\tau)} .
 \label{2nd}
\end{eqnarray}
By interchanging the arguments $t$ and $\tau$ of $\tilde{H}^L_I$ in Eq. (\ref{2nd}), one obtains the expression for the 
third term:\\
\begin{eqnarray}
 &&\sum^{\prime}_{n}\langle 10|{~_{ph}}\langle n| \tilde{H}_I^L(\tau)|0\rangle_{ph}\tilde{\rho}_s(t) 
 {_{ph}}\langle 0 | \tilde{H}_I^L(t) |n\rangle_{ph}|01\rangle \nonumber \\
 &&=\kappa^2 \sum^{\prime}_{n} C_{n} \bigg[\kappa^2\langle 10|\tilde{\rho}_s(t) |01\rangle  \Big\{(-1)^{(n_1+n_2)}q(\tau)q(t)p(\tau)p(t)
 -q(\tau)p^* (t) p(\tau)q(t) \nonumber \\
 &&~~~~~~~~~~~~~~~~~~~~~~~~~~~~~~~~~~~~~~-p^*(\tau)q(t)q(\tau)p(t)+ (-1)^{(n_1+n_2)} p^* (\tau)p^* (t)q(\tau)q(t)\Big\}\nonumber \\
&& ~~~~~~~~~~~~~~~~+\langle 01|\tilde{\rho}_s(t) |10\rangle \Big\{(-1)^{(n_1+n_2)}p^2(\tau)p^2(t)+\kappa^2 p^2(\tau) q^2 (t)\nonumber \\
 &&~~~~~~~~~~~~~~~~~~~~~~~~~~~~~~~~~~~~~~+\kappa^2 q^2(\tau) p^2 (t)+(-1)^{(n_1+n_2)}\kappa^4 q^2(\tau) q^2 (t)\Big\}\nonumber \\
 &&~~~~~~~~~~~~~~~~+i\kappa\langle 10|\tilde{\rho}_s(t) |10\rangle  \Big\{-(-1)^{(n_1+n_2)}q(\tau)p^2(t)p(\tau)-\kappa^2 q(\tau)q^2(t)p(\tau)\nonumber\\
 &&~~~~~~~~~~~~~~~~~~~~~~~~~~~~~~~~~~~~~+p^*(\tau)p^2(t)q(\tau)+(-1)^{(n_1+n_2)}\kappa^2 p^*(\tau)q^2(t)q(\tau) \Big \}\nonumber \\
 &&~~~~~~~~~~~~~~~~+i\kappa\langle 01|\tilde{\rho}_s(t) |01\rangle \Big\{(-1)^{(n_1+n_2)}p^2(\tau)q(t)p(t)-p^*(t)p^2(\tau) q(t) \nonumber \\
 &&~~~~~~~~~~~~~~~~~~~~~~~~~~~~~~~~~~~~~+\kappa^2 q^2(\tau)q(t)p(t)-(-1)^{(n_1+n_2)}\kappa^2q^2 (\tau) p^*(t)q(t)\Big\}\Bigg]e^{-i\omega_n (t-\tau)} .
 \label{3rd}
 \end{eqnarray}

Lastly,  we get the following fourth term:\\
\begin{eqnarray}
&&\sum^{\prime}_{n}\langle 10|\tilde{\rho}_s(t)  _{ph}\langle 0|
 \tilde{H}_I^L(\tau)| n \rangle_{ph}~ _{ph}\langle n | \tilde{H}_I^L(t)| 0\rangle_{ph}|01\rangle  \nonumber \\
&&=\kappa^2 \sum^{\prime}_{n} C_{n} \bigg[ \langle 10|\tilde{\rho}_s(t) |01\rangle \Big\{p(t)p(t-\tau)p^*(\tau)+
{(-1)^{(n_1+n_2)}}\kappa^2 q(t)q(t-\tau)
p^*(\tau) \nonumber \\
&&~~~~~~~~~~~~~~~~~~~~~~~~~~~~~~~~~~~-\kappa^2(-1)^{(n_1+n_2)} p(t)q(t-\tau)q(\tau) +\kappa^2  q(t)p^*(t-\tau)q(\tau)\Big\} \nonumber \\
&&~~~~~~~~~~~~~~~~+i\kappa \langle 10|\tilde{\rho}_s(t) |10\rangle \Big\{p(t)p(t-\tau)q(\tau)+(-1)^{(n_1+n_2)}\kappa^2 q(t)q(t-\tau)q(\tau)\nonumber \\
&&~~~~~~~~~~~~~~~~~~~~~~~~~~~~~~~~~~~~~ +(-1)^{(n_1+n_2)}p(t)q(t-\tau)p(\tau)-q(t)p^* (t-\tau) p(\tau)\Big\} \bigg]e^{i\omega_n (t-\tau)}.
\label{4th}
\end{eqnarray}
When $|\kappa| \ll |\Delta\varepsilon|$ one can write [using the definition of $q(t)$ in the main text]
\begin{eqnarray}
\kappa q(t)\approx 2\frac{\kappa}{\Delta\varepsilon}{\rm sin}(\frac{\Delta\varepsilon}{2}t) \ll 1.
\end{eqnarray}
So, for all practical purposes, one can neglect terms involving $\kappa q$ in Eqs. (\ref{1st})--(\ref{4th}). Moreover, based on 
Eqs. (\ref{basis1}) and (\ref{basis2}), we can write 
\begin{eqnarray}
 p(t)\approx e^{i\frac{\Delta\varepsilon}{2}t} .
\end{eqnarray}
By implementing the above approximations, one can write  Eqs. (\ref{1st})--(\ref{4th}) as
\begin{eqnarray}
\sum^{\prime}_{n} \langle 10|_{ph}\langle 0 |
 \tilde{H}_I^L(t)| n \rangle_{ph}~ _{ph}\langle n| \tilde{H}_I^L(\tau)| 0 \rangle_{ph} \tilde{\rho}_s(t) |01\rangle 
\approx \kappa^2 \sum^{\prime}_{n} C_n \langle 10|\tilde{\rho}_s(t)|01\rangle e^{i\Delta\varepsilon(t-\tau)}  e^{-i\omega_n (t-\tau)} ,
 \label{1stap}
\end{eqnarray}

\begin{eqnarray}
\sum^{\prime}_{n}\langle 10|{~_{ph}}\langle n| \tilde{H}_I^L(t)|0\rangle_{ph}\tilde{\rho}_s(t) 
 {_{ph}}\langle 0 | \tilde{H}_I^L(\tau) |n\rangle_{ph}|01\rangle 
\approx \kappa^2 \sum^{\prime}_{n} C_n \langle 01|\tilde{\rho}_s(t)|10\rangle (-1)^{(n_1+n_2)} e^{i\Delta\varepsilon(t+\tau)}e^{i\omega_n (t-\tau)} ,
 \label{2ndap}
\end{eqnarray}

\begin{eqnarray}
\sum^{\prime}_{n}\langle 10|{~_{ph}}\langle n| \tilde{H}_I^L(\tau)|0\rangle_{ph}\tilde{\rho}_s(t) 
 {_{ph}}\langle 0 | \tilde{H}_I^L(t) |n\rangle_{ph}|01\rangle 
\approx \kappa^2 \sum^{\prime}_{n} C_n \langle 01|\tilde{\rho}_s(t)|10\rangle (-1)^{(n_1+n_2)} e^{i\Delta\varepsilon(t+\tau)}e^{-i\omega_n (t-\tau)}, 
\end{eqnarray}
and
\begin{eqnarray}
\sum^{\prime}_{n}\langle 10|\tilde{\rho}_s(t)  _{ph}\langle 0|
 \tilde{H}_I^L(\tau)| n \rangle_{ph}~ _{ph}\langle n | \tilde{H}_I^L(t)| 0\rangle_{ph}|01\rangle 
\approx \kappa^2 \sum^{\prime}_{n} C_n \langle 10|\tilde{\rho}_s(t)|01\rangle e^{i\Delta\varepsilon(t-\tau)}  e^{i\omega_n (t-\tau)} ,
\end{eqnarray}
respectively. 

Next, the differential equation for the excited state population $\langle 10|\rho_s(t)|10\rangle$ is written as 
\begin{eqnarray}
 \frac{d \langle 10| \tilde{\rho}_s(t)|10\rangle}{dt} &=& - \sum_{n} \int_0^t d\tau \bigg [~ \langle 10|_{ph}\langle 0 |
 \tilde{H}_I^L(t)| n \rangle_{ph}~ _{ph}\langle n| \tilde{H}_I^L(\tau)| 0 \rangle_{ph} \tilde{\rho}_s(t) |10\rangle
\nonumber \\ 
 &&~~~~~~~~~~~~~~~~~~~ - \langle 10|{~_{ph}}\langle n| \tilde{H}_I^L(t)|0\rangle_{ph}\tilde{\rho}_s(t) {_{ph}}\langle 0 | \tilde{H}_I^L(\tau) |n\rangle_{ph}|10\rangle \nonumber \\
 && ~~~~~~~~~~~~~~~~~~~- \langle 10|{~_{ph}}\langle n| \tilde{H}_I^L(\tau)| 0\rangle_{ph} \tilde{\rho}_s(t) {_{ph}}\langle 0| \tilde{H}_I^L(t) |n\rangle_{ph}|10\rangle 
 \nonumber \\
&& ~~~~~~~~~~~~~~~~~~~
+ \langle 10|\tilde{\rho}_s(t)  _{ph}\langle 0|
 \tilde{H}_I^L(\tau)| n \rangle_{ph}~ _{ph}\langle n | \tilde{H}_I^L(t)| 0\rangle_{ph}|10\rangle  
 \bigg ] .
 \label{masdia}
\end{eqnarray}
In the integrand of the above equation, we express  the first term as\\
\begin{eqnarray}
  &&\sum^{\prime}_{n} \langle 10|_{ph}\langle 0 |
 \tilde{H}_I^L(t)| n \rangle_{ph}~ _{ph}\langle n| \tilde{H}_I^L(\tau)| 0 \rangle_{ph} \tilde{\rho}_s(t) |10\rangle \nonumber \\
 &&=\kappa^2 \sum^{\prime}_{n} C_n \bigg[ \langle 10|\tilde{\rho}_s(t) |10\rangle \Big\{p(t)p(t-\tau)p^*(\tau)+(-1)^{(n_1+n_2)} \kappa^2 q(t)q(t-\tau)
p^*(\tau)\nonumber \\
&&~~~~~~~~~~~~~~~~~~~~~~~~~~~~~~~~~~~+ \kappa^2 q(t) p^*(t-\tau)q(\tau)-(-1)^{(n_1+n_2)}\kappa^2 p(t)q(t-\tau)q(\tau) \Big\} \nonumber \\
&&~~~~~~~~~~~~~~~~-i\kappa \langle 01|\tilde{\rho}_s(t) |10\rangle \Big\{p(t)p(t-\tau)q(\tau)+\kappa^2 (-1)^{(n_1+n_2)}q(t)q(t-\tau)q(\tau) \nonumber \\
&&~~~~~~~~~~~~~~~~~~~~~~~~~~~~~~~~~~~~~ -q(t)p^* (t-\tau) p(\tau)+(-1)^{(n_1+n_2)} p(t)q(t-\tau)p(\tau)\Big\}\bigg] e^{-i\omega_n (t-\tau)} ,
\label{1stdia}
\end{eqnarray}
the second term as
\begin{eqnarray}
 &&\sum^{\prime}_{n}\langle 10|{~_{ph}}\langle n| \tilde{H}_I^L(t)|0\rangle_{ph}\tilde{\rho}_s(t) 
 {_{ph}}\langle 0 | \tilde{H}_I^L(\tau) |n\rangle_{ph}|10\rangle \nonumber \\
 &&=\kappa^2 \sum^{\prime}_{n} C_n \bigg[\kappa^2 \langle 10|\tilde{\rho}_s(t)|10\rangle  \Big\{p(t)p^*(\tau)q(t)q(\tau) 
 -(-1)^{(n_1+n_2)}p(t)q(\tau)q(t)p(\tau) \nonumber \\
 &&~~~~~~~~~~~~~~~~~~~~~~~~~~~~~~~~~~~ -(-1)^{(n_1+n_2)}q(t)p^*(\tau)p^*(t)q(\tau)+q(t)q(\tau)p^*(t)p(\tau)\Big\}\nonumber \\
 &&~~~~~~~~~~~~~~~+\langle 01|\tilde{\rho}_s(t)|01\rangle \Big\{ p^2(t)p^{*}~^{2}(\tau)+(-1)^{(n_1+n_2)}\kappa^2p^2(t)q^2(\tau)\nonumber \\
 &&~~~~~~~~~~~~~~~~~~~~~~~~~~~~~~~~~~+(-1)^{(n_1+n_2)}\kappa^2  q^2(t)p^{*}~^2(\tau)+\kappa^4 q^2(t)q^2(\tau)\Big\}\nonumber \\
 &&~~~~~~~~~~~~~~~+i\kappa\langle 10|\tilde{\rho}_s(t)|01\rangle \Big\{-p(t)p^{*}~^2(\tau)q(t)-(-1)^{(n_1+n_2)}\kappa^2p(t)q^2(\tau)q(t)\nonumber \\
 &&~~~~~~~~~~~~~~~~~~~~~~~~~~~~~~~~~~~~+(-1)^{(n_1+n_2)}q(t)p^{*}~^2(\tau)p^*(t)+\kappa^2q(t)q^2(\tau)p^*(t)\Big\} \nonumber \\
 &&~~~~~~~~~~~~~~~+i\kappa\langle 01|\tilde{\rho}_s(t)|10\rangle  \Big\{p^2(t)p^*(\tau)q(\tau)-(-1)^{(n_1+n_2)}p^2(t)q(\tau)p(\tau)\nonumber \\
 &&~~~~~~~~~~~~~~~~~~~~~~~~~~~~~~~~~~~~+(-1)^{(n_1+n_2)}\kappa^2 q^2(t)p^*(\tau)q(\tau)-\kappa^2q^2(t)q(\tau)p(\tau)\Big\}\bigg]e^{i\omega_n (t-\tau)} ,
 \label{2nddia}
\end{eqnarray}
the third term as
\begin{eqnarray}
  &&\sum^{\prime}_{n}\langle 10|{~_{ph}}\langle n| \tilde{H}_I^L(\tau)|0\rangle_{ph}\tilde{\rho}_s(t) 
 {_{ph}}\langle 0 | \tilde{H}_I^L(t) |n\rangle_{ph}|10\rangle \nonumber \\
 &&=\kappa^2 \sum^{\prime}_{n} C_n \bigg[\kappa^2 \langle 10|\tilde{\rho}_s(t)|10\rangle  \Big\{p(\tau)p^*(t)q(\tau)q(t) 
 -(-1)^{(n_1+n_2)}p(\tau)q(t)q(\tau)p(t)\nonumber \\
 &&~~~~~~~~~~~~~~~~~~~~~~~~~~~~~~~~~~~ -(-1)^{(n_1+n_2)}q(\tau)p^*(t)p^*(\tau)q(t)+q(\tau)q(t)p^*(\tau)p(t)\Big\}\nonumber \\
 &&~~~~~~~~~~~~~~~+\langle 01|\tilde{\rho}_s(t)|01\rangle \Big\{ p^2(\tau)p^{*}~^{2}(t)+(-1)^{(n_1+n_2)}\kappa^2p^2(\tau)q^2(t)\nonumber \\
 &&~~~~~~~~~~~~~~~~~~~~~~~~~~~~~~~~~~+(-1)^{(n_1+n_2)}\kappa^2  q^2(\tau)p^{*}~^2(t)+\kappa^4 q^2(\tau)q^2(t)\Big\}\nonumber \\
 &&~~~~~~~~~~~~~~~+i\kappa\langle 10|\tilde{\rho}_s(t)|01\rangle \Big\{-p(\tau)p^{*}~^2(t)q(\tau)-(-1)^{(n_1+n_2)}\kappa^2p(\tau)q^2(t)q(\tau)\nonumber \\
 &&~~~~~~~~~~~~~~~~~~~~~~~~~~~~~~~~~~~~ +(-1)^{(n_1+n_2)}q(\tau)p^{*}~^2(t)p^*(\tau)+\kappa^2q(\tau)q^2(t)p^*(\tau)\Big\} \nonumber \\
 &&~~~~~~~~~~~~~~~+i\kappa\langle 01|\tilde{\rho}_s(t)|10\rangle  \Big\{p^2(\tau)p^*(t)q(t)-(-1)^{(n_1+n_2)}p^2(\tau)q(t)p(t)\nonumber \\
 &&~~~~~~~~~~~~~~~~~~~~~~~~~~~~~~~~~~~~ +(-1)^{(n_1+n_2)}\kappa^2 q^2(\tau)p^*(t)q(t)-\kappa^2q^2(\tau)q(t)p(t)\Big\}\bigg]e^{-i\omega_n (t-\tau)} ,
 \label{3rddia}
\end{eqnarray}
and the fourth term as
\begin{eqnarray}
 &&\sum^{\prime}_n\langle 10|\tilde{\rho}_s(t)  _{ph}\langle 0|
 \tilde{H}_I^L(\tau)| n \rangle_{ph}~ _{ph}\langle n | \tilde{H}_I^L(t)| 0\rangle_{ph}|10\rangle \nonumber \\
 &&=\kappa^2 \sum^{\prime}_{n} C_n \bigg[ \langle 10|\tilde{\rho}_s(t) |10\rangle \Big\{p^*(t)p^*(t-\tau)p(\tau)+(-1)^{(n_1+n_2)} \kappa^2 q(t)q(t-\tau)
p(\tau)\nonumber \\
&&~~~~~~~~~~~~~~~~~~~~~~~~~~~~~~~~~~~-(-1)^{(n_1+n_2)}\kappa^2 q(\tau)q(t-\tau)p^*(t)+\kappa^2 q(t)p(t-\tau)q(\tau)\Big\}\nonumber \\
&&~~~~~~~~~~~~~~~~+i\kappa\langle 10|\tilde{\rho}_s(t) |01\rangle  \Big\{p^*(t)p^*(t-\tau)q(\tau)+(-1)^{(n_1+n_2)}\kappa^2 q(t)q(t-\tau)q(\tau)\nonumber\\
&&~~~~~~~~~~~~~~~~~~~~~~~~~~~~~~~~~~~~~ +(-1)^{(n_1+n_2)} q(t-\tau)p^*(t)p^*(\tau)-q(t)p(t-\tau) p^*(\tau)\Big\}\bigg]e^{i\omega_n (t-\tau)} .
\label{4thdia}
\end{eqnarray}
When $|\kappa| \ll |\Delta\varepsilon|$, we can simplify Eqs. (\ref{1stdia})--(\ref{4thdia}) as
\begin{eqnarray}
\sum^{\prime}_{n} \langle 10|_{ph}\langle 0 |
 \tilde{H}_I^L(t)| n \rangle_{ph}~ _{ph}\langle n| \tilde{H}_I^L(\tau)| 0 \rangle_{ph} \tilde{\rho}_s(t) |10\rangle
\approx\kappa^2 \sum^{\prime}_{n} C_n \langle 10|\tilde{\rho}_s(t) |10\rangle e^{i\Delta\varepsilon(t-\tau)}  e^{-i\omega_n (t-\tau)},
\end{eqnarray}
\begin{eqnarray}
\sum^{\prime}_{n}\langle 10|{~_{ph}}\langle n| \tilde{H}_I^L(t)|0\rangle_{ph}\tilde{\rho}_s(t) 
 {_{ph}}\langle 0 | \tilde{H}_I^L(\tau) |n\rangle_{ph}|10\rangle 
\approx \kappa^2 \sum^{\prime}_{n} C_n \Big (1-\langle 10|\tilde{\rho}_s(t) |10\rangle \Big ) e^{i\Delta\varepsilon(t-\tau)}  e^{i\omega_n (t-\tau)} ,
\end{eqnarray}
\begin{eqnarray}
\sum^{\prime}_{n}\langle 10|{~_{ph}}\langle n| \tilde{H}_I^L(\tau)|0\rangle_{ph}\tilde{\rho}_s(t) 
 {_{ph}}\langle 0 | \tilde{H}_I^L(t) |n\rangle_{ph}|10\rangle 
\approx \kappa^2 \sum^{\prime}_{n} C_n \Big (1-\langle 10|\tilde{\rho}_s(t) |10\rangle \Big) e^{-i\Delta\varepsilon(t-\tau)}  e^{-i\omega_n (t-\tau)} ,
\end{eqnarray}
and
\begin{eqnarray}
\sum^{\prime}_n\langle 10|\tilde{\rho}_s(t)  _{ph}\langle 0|
 \tilde{H}_I^L(\tau)| n \rangle_{ph}~ _{ph}\langle n | \tilde{H}_I^L(t)| 0\rangle_{ph}|10\rangle 
\approx \kappa^2 \sum^{\prime}_{n} C_n \langle 10|\tilde{\rho}_s(t) |10\rangle e^{-i\Delta\varepsilon(t-\tau)}  e^{i\omega_n (t-\tau)} ,
\end{eqnarray}
respectively. The multimode case can be derived in a similar fashion by replacing $\omega_n$, $C_n$, $\kappa$, and $n_i$  with 
$\bar{\omega}_n$, $\bar{C}_n$, $\bar{\kappa}$, and $\sum_k n^k_i$, respectively.
\end{widetext}

\pagebreak
\section{Popular Summary}

Photosynthesis is a process used by plants and other organisms to convert light
into the chemical energy used by most life on earth. Understanding the highly
efficient transport of absorbed light-energy through molecules in photosynthesis
is of significant scientific interest and also key to designing light-harvesting
technology.  Surprisingly, experiments reveal that superposition principle
of quantum mechanics plays a crucial role in this biological transport process;
it is the interference of the superposed states (much like the interference in waves)
and its preservation during transport that is key to the high efficiency of the process.  
However, the precise features that preserve the quantum interference during
transport  have remained elusive.  Next, also of immense interest is the successful
exploitation of quantum mechanics for quantum computation so as to solve complex
problems that are intractable by present-day classical computers.  A qubit is the
building block of a quantum computer much like a bit in a classical computer.
Superposition principle is what distinguishes a quantum bit (qubit) from a
classical bit. Although a few promising candidates for a qubit exist, none of
them preserve superposition of the two states of a qubit for extended time periods.
Unfortunately, interaction of the excited molecules (in photosynthesis) and qubits
(in a quantum computer) with the ubiquitous environmental noise is inevitable;
this coupling to noise can degrade the superposition of states in a system of molecules
or a qubit, i.e., can produce decoherence in them.  Our work identifies a few key
features that reduce decoherence significantly.

      The excitations in photosynthesis and the qubit  in quantum computation
can be modeled by a spin or a hard-core boson (HCB), i.e., a boson that does not
allow another one to occupy the same site.  Crucially, the system and the environment
(comprising of optical phonons) should be initially uncorrelated in the frame of
reference where the HCB is dressed by the deformation it produces in the lattice
environment. Our work shows that coherence is maximized if HCBs have distance-independent
hopping in a lattice.  Furthermore, if each site in the lattice has a different potential,
then coherence is preserved when the potential differences between sites is as far
away as possible from any of the environmental eigenenergies (and particularly those
close to twice the lattice deformation energy). Interestingly, we show that the stronger
the HCB couples with the environment, the lesser is the decoherence. 

     Thus, our work would help in developing synthetic light-absorbing materials with
efficient energy transfer as well as in realizing  charge qubits with small decoherence.
An oxide- (i.e., manganite-) based double quantum dot can serve as a robust charge qubit
and meet the demands put by miniaturization on computer technology.


\begin{thebibliography}{47}
\bibitem{sm_par}
For a large $N$ system, the largest energy contribution due to terms involving closed loops is intensive. The dominant
contribution, in second and higher order perturbation,
comes from open loops.
 The approximate open loop terms in zeroth, second,
and third order perturbation are, respectively, $J_{\perp}e^{-g^2} \sum_{i,j\neq i} b^{\dagger}_i b_j$, $[NJ^2_{\perp} e^{-g^2}/(g^2 \omega)]
\sum_{i,j\neq i} b^{\dagger}_i b_j$,
and $[N^2 J^3_{\perp} e^{-g^2}/(g^2 \omega)^2 ]\sum_{i,j \neq i} b^{\dagger}_i b_j$; then, the small parameter
is $J^{\star}_{\perp}/(g^2 \omega)$.
\end{thebibliography}
\end{document}